\documentclass{elsart}
\usepackage{natbib}
\usepackage{graphicx}
\begin{document}
\runauthor{Wanajo and Ishimaru}
\begin{frontmatter}
\title{{\boldmath $r$}-Process Calculations and Galactic Chemical Evolution
}
\author[Tokyo]{Shinya Wanajo}
\and
\author[Kogakuin]{Yuhri Ishimaru}

\address[Tokyo]{Research Center for the Early Universe,
   Graduate School of Science, University of Tokyo,
   Bunkyo-ku, Tokyo 113-8654, Japan;
   wanajo@resceu.s.u-tokyo.ac.jp}
\address[Kogakuin]{Academic Support Center, Kogakuin University,
   Hachioji, Tokyo 192-0015, Japan;
   kt13121@ns.kogakuin.ac.jp}

\bigskip

\address{Nuclear
      Physics A (Special Issue on Nuclear Astrophysics /
      eds. K. Langanke, F.-K. Thielemann, \& M. Wiescher), in press}

\begin{abstract}
While the origin of $r$-process nuclei remains a long-standing mystery,
recent spectroscopic studies of extremely metal-poor stars in the
Galactic halo strongly suggest that it is associated with core-collapse
supernovae. In this article, an overview of the recent theoretical
studies of the $r$-process is presented with a special emphasis on the
astrophysical scenarios related to core-collapse supernovae. We also
review a recent progress of the Galactic chemical evolution studies as
well as of the spectroscopic studies of extremely metal-poor halo stars,
which provide us important clues to better understanding of the
astrophysical $r$-process site.
\end{abstract}
\end{frontmatter}

\section{Introduction}

The rapid neutron-capture process ($r$-process) accounts for the
production of about half of nuclei heavier than iron, such as the bulk
of noble metals (e.g., silver, platinum, and gold) and all actinides
(e.g., thorium, uranium, and plutonium). While the basic picture of the
$r$-process, as well as of the slow neutron-capture process
($s$-process), from the nuclear physics point of view is well
established about a half century ago \cite{Burb57, Came57}, its
astrophysical origin has been still unknown. In the last decade, many
theoretical efforts have been dedicated to the studies related to the
``neutrino wind'' scenario, i.e., the $r$-process is expected to take
place in the high-entropy, neutrino-heated ejecta from the nascent
neutron star (NS) in a core-collapse supernova \cite[i.e., Type~II/Ibc
SNe,][]{Woos92, Meye92, Woos94, Taka94, Qian96, Card97, Otsu00, Sumi00,
Wana01, Thom01}. A few other scenarios have been also suggested, which
include the ``prompt explosion'' from a low mass SN \cite{Sumi01,%
Wana03}, the ``NS merger'' \cite{Frei99b, Gori05a}, and the ``collapsar''
from a massive progenitor \cite{MacF99, Prue04}. All the scenarios
proposed above involve, however, severe problems that remain to be
solved, and no consensus has yet been achieved.

Despite difficulties in theoretical studies, recent comprehensive
spectroscopic analyses of extremely metal-poor stars in the Galactic
halo, aided with Galactic chemical evolution studies, have provided us
important clues to the astrophysical origin of $r$-process nuclei. In
particular, discoveries of extremely metal-poor, $r$-process-enhanced
stars with remarkable agreement of their abundance patterns to the
scaled solar $r$-process curve strongly support the idea that the
$r$-process nuclei originate from short-lived massive stars, i.e.,
core-collapse SNe \cite{Sned96, Cayr01, Hill02, Sned03}. Furthermore,
the observed large star-to-star scatters of $r$-process elements with
respect to iron suggest that the progenitors responsible for the
$r$-process abundance production is limited to a small mass range, when
combined with Galactic chemical evolution models \cite{Ishi99, Trav99,%
Tsuj00, Arga04}.

In the subsequent sections, an overview of the current status of
explorations of the astrophysical $r$-process origin is presented from
different points of views, i.e., nucleosynthesis studies related to, in
particular, core-collapse SNe (\S~1), and chemical evolution studies of
the Galactic halo, along with recent spectroscopic analyses of extremely
metal-poor stars (\S~2). Conclusions follow (\S~3).

\section{{\boldmath $r$}-Process Calculations}

The $r$-process proceeds through the neutron-rich region far from
$\beta$-stability in the nuclide chart (Fig.~1), which needs a high
neutron-to-seed abundance ratio ($\ge 100$, where ``seed'' is the heavy
nuclei with $A\sim 60-90$) at the beginning of the $r$-process phase
($T_9 \sim 3$, where $T_9 \equiv T/10^9 {\rm ~K}$ and $T$ is
temperature). The requirements on the physical conditions here are
threefold -- low electron fraction ($Y_e$, number of protons per
baryon), high entropy ($S \propto T^3/\rho$ in radiation dominated
matter, where $\rho$ is mass density), and short dynamic timescale
(e.g., $\tau_{\rm dyn} = |\rho/(d\rho/dt)|_{T = 0.5 {\rm MeV}}$). Matter
with $Y_e < 0.5$ is called {\it neutron rich}. In particular, the matter
with $Y_e < 0.3$ contains free neutrons even in the nuclear statistical
equilibrium (NSE) at relatively low temperature. Hence, the matter with
significantly low $Y_e$, say, $< 0.2$, may naturally lead to robust
$r$-processing, less dependent on $S$ and $\tau_{\rm dyn}$. Even if the
matter is only moderately neutron-rich, say, $Y_e \sim 0.4$,
sufficiently high $S$ ($ > {\rm a~few~} 100 N_A k$) or short $\tau_{\rm
dyn}$ ($<$ a few 10 ms) inhibit free nucleons and $\alpha$-particles to
assemble to heavier nuclei during the $\alpha$-process phase ($T_9
\approx 7 - 3$), and may leave sufficient free neutrons needed for a
successful $r$-process. Our main goal is to find such physical
conditions by (to some extent) realistic astrophysical modelings.

\begin{figure}[t]
\begin{center}
\includegraphics*[width=\textwidth]{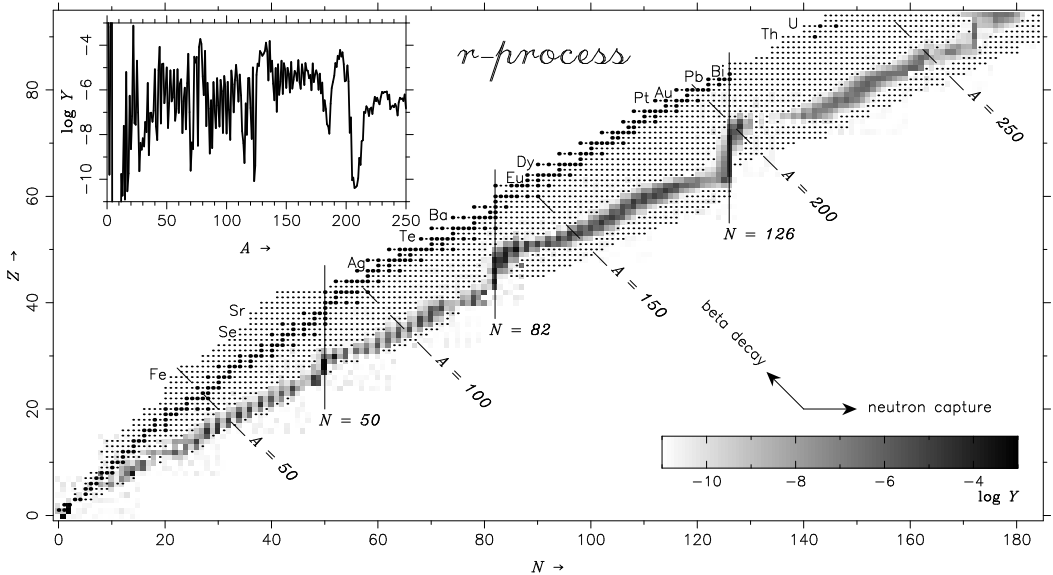}
\end{center}
\caption{Snapshot of the nucleosynthesis calculation at the end of the
$r$-process phase. The abundances are shown by the grey image in the
nuclide chart.  The abundance curve as a function of mass number is
shown in the upper left. The nuclei included in the reaction network are
denoted by dots, with the stable and long-lived isotopes represented by
large dots \cite{Wana02}.}
\end{figure}

\subsection{Neutrino Wind}

The bottleneck reaction from light ($Z < 6$) to heavy ($Z\ge 6$) nuclei
in the neutron-rich environment is the three-body interaction, $\alpha
(\alpha n, \gamma)^9$Be followed by $^9$Be$(\alpha, n)^{12}$C, whose
rate is proportional to $\rho^2$. This is why higher $S$ (i.e., lower
$\rho$ in the radiative condition) is favored for $r$-process, which
tends to leave more free neutrons. Such high entropy is expected to
realize in the neutrino-heated ejecta (``neutrino wind'') from the
nascent NS in a core-collapse SN, and many efforts have been devoted to
the study of this scenario in the last decade. Woosley et
al. \cite{Woos94} have demonstrated that an excellent fit to the solar
$r$-process abundance curve is obtained in their nucleosynthesis
calculations with the thermodynamic trajectories from the hydrodynamic
simulation of a $20 M_\odot$ ``delayed'' SN explosion. The high entropy
($\sim 400 N_A k$) that led to a successful $r$-processing was not,
however, duplicated by subsequent theoretical studies \cite{Taka94,%
Qian96}. In the following, the current status of the theoretical studies
of the neutrino wind is described based on our recent works
\cite{Otsu00, Wana01, Wana02} that result in similar conclusions to
other studies \cite{Qian96, Card97, Sumi00, Thom01}.

\subsubsection{Wind Properties and Nucleosynthesis}

After several 100~ms from the core bounce, the hot convective bubbles
are evacuated from the proto-NS surface, and the winds driven by
neutrino heating emerge, as can be seen in some hydrodynamic simulations
of ``successful'' delayed SN explosions \cite[e.g.,][]{Woos94,
Bura05}. During this wind phase, a steady flow approximation may be
justified. Assuming the spherical symmetry, the equations of baryon,
momentum, and mass-energy conservation with the Schwarzschild metric
\cite[e.g., equations~(1)-(3) in][]{Wana01} can be solved
numerically. Thus, once the NS mass ($M$), the neutrino sphere radius
($R_\nu$), and the neutrino luminosity ($L_\nu$) are specified along
with the mass ejection rate ($\dot M$) as the boundary condition, the
wind solution can be obtained.

Figs.~2a-c show the maximum mass ejection rate ($\dot M_{\rm max}$,
i.e., for transonic solutions), the entropy per baryon ($S/k$ at
$T=0.5$~MeV), and the timescale ($\tau$, as the time for material to
cool from $T=0.5$ to 0.2~MeV as a measure of the duration of the seed
abundance production), respectively, as functions of $L_\nu$ that is
assumed to be equal for all neutrino flavors \cite{Wana01}. The results
are compared to those with post-Newtonian corrections \cite{Qian96} and
with fully general relativistic hydrodynamic calculations \cite{Sumi00},
which are in good agreement each other. As can be seen, the entropy is
$\sim 120 N_A k$ at $L_\nu = 10^{51}$~erg~s$^{-1}$ for $M = 1.4 M_\odot$
and $R_\nu = 10$~km (model~A, dot-dashed line), which is more than three
times smaller than that in Woosley et al. \cite{Woos94}. However, the
entropy can be $\sim 200 N_A k$ for a very {\it compact} proto-NS, i.e.,
$M = 1.4 M_\odot$ and $R_\nu = 7$~km (model~B, dashed line) or, $M = 2.0
M_\odot$ and $R_\nu = 10$~km (model~C, solid line), where the general
relativistic effects are of particular importance. Not only to the
entropy, the general relativity helps to reduce $\tau$ as can be seen in
Fig.~2c.

\begin{figure}[t]
\begin{minipage}[t]{68mm}
\includegraphics*[width=\textwidth]{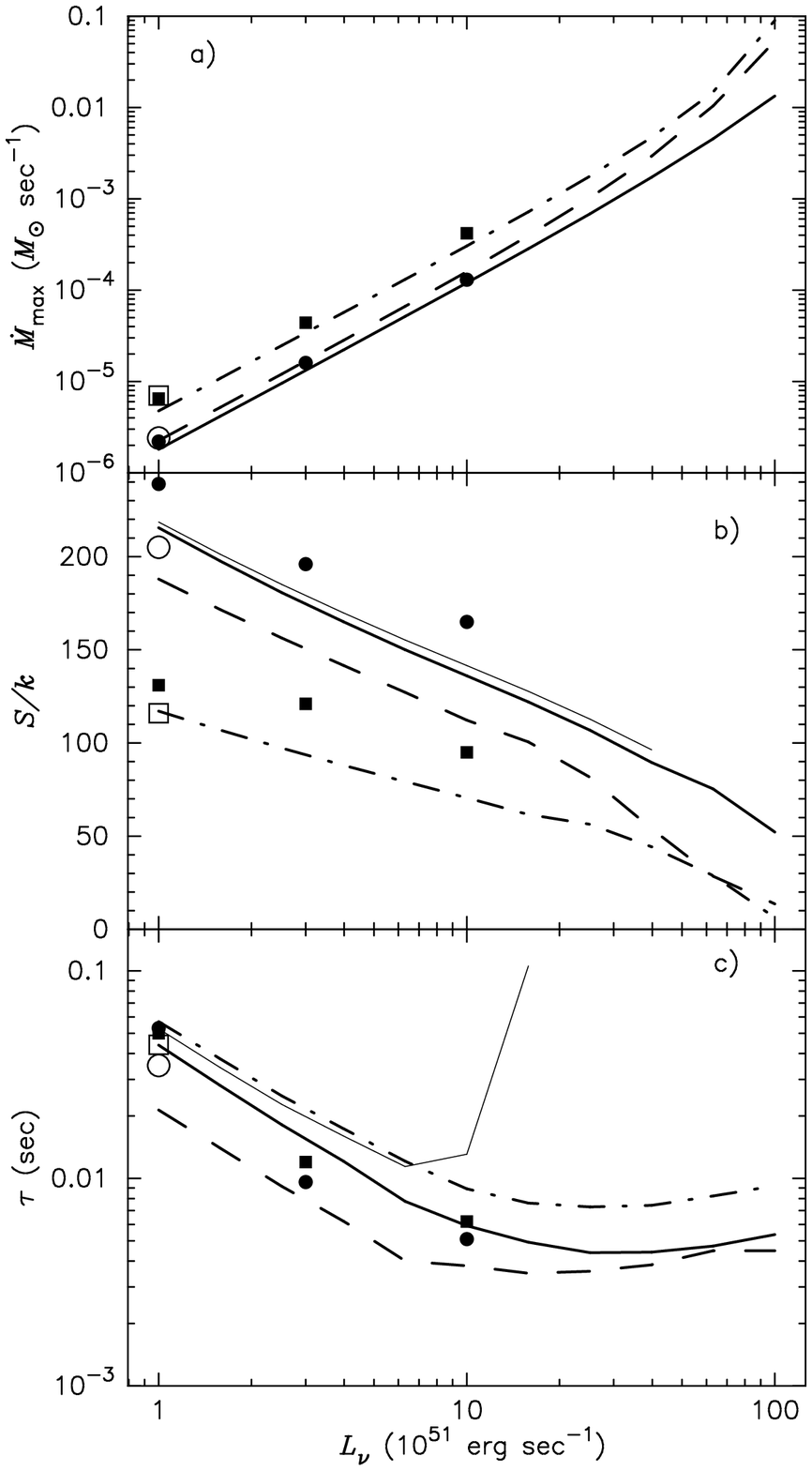} \caption{(a) The maximum
mass ejection rates, (b) entropies, and (c) timescales for models~A
(dot-dashed line), B (dashed line), and C (thick-solid line), as
functions of $L_\nu$ for the transonic winds. The thin-solid lines are
for the subsonic wind with $\dot M = 0.995 \times \dot M_{\rm max}$ for
model~C. Also denoted are the results from \cite{Sumi00} (filled squares
and circles) and \cite{Qian96} (open squares and circles). The squares
and circles are the results with the same model parameters $M$ and
$R_\nu$ as the models A and B, respectively.}
\end{minipage}
\hspace{\fill}
\begin{minipage}[t]{68mm}
\includegraphics*[width=\textwidth]{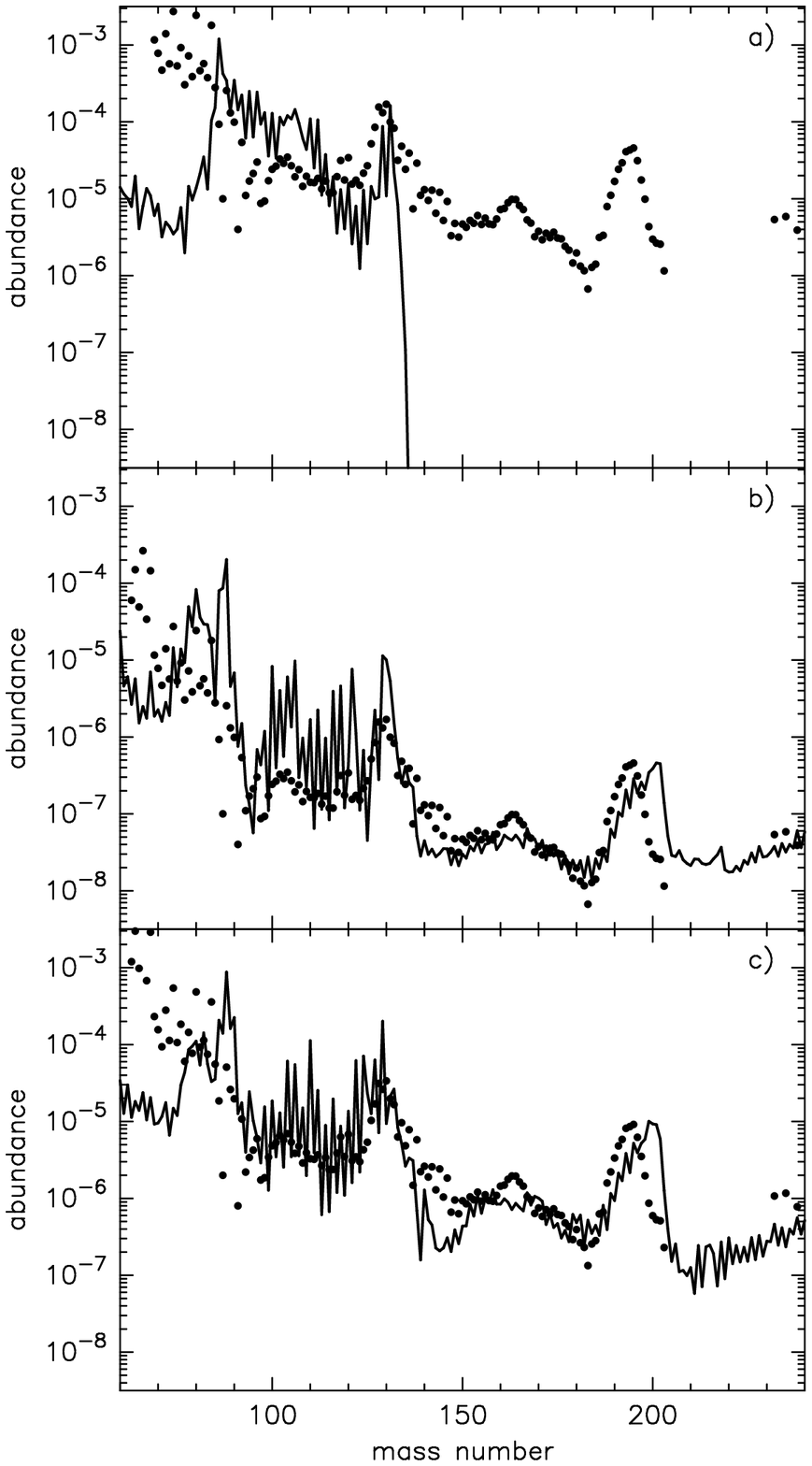} \caption{The mass-weighted
integrated yields for models~A (a), B (b), and C (c) as functions of
mass number (lines). Also denoted are the scaled solar $r$-process
abundances \cite{Kapp89} (points), which are scaled to match the heights
of the second (a) and third (b and c) $r$-process peaks,
respectively. For model~A, only the nuclei between the first and second
$r$-process peaks are produced. The third peak is formed for models~B
and C. Note a significant overproduction of nuclei near $A = 90$ for
both (b) and (c).}
\end{minipage}
\end{figure}

The yields of $r$-process nuclei are obtained by application of an
extensive nuclear reaction network that consists of $\sim 5000$ species
along with all relevant nuclear reaction and weak rates \cite[Fig.~1,
see][for the nuclear data inputs]{Wana01, Wana02, Wana03, Wana04}. The
results for the above three models are shown in Fig.~3, where $Y_e$ is
taken to be 0.4 and the abundances for constant $L_\nu$'s are
mass-averaged assuming a time evolution of $L_\nu$ from $4 \times
10^{52}$ to $1 \times 10^{51}$~erg~s$^{-1}$ \cite[see][for more
detail]{Wana01, Wana02}. For a {\it typical} proto-NS (model~A), only
the nuclei between the first ($A = 80$) and second ($A = 130$)
$r$-process peaks are produced, owing to its insufficient entropy. In
contrast, for very {\it compact} proto-NSs (models~B and C), the third
$r$-process peak ($A = 195$) forms and each abundance curve is in
reasonable agreement with the solar $r$-process abundance
distribution. This is not only due to the moderately high entropy ($\sim
100-200 N_A k$), which is still a half that in Woosley et
al. \cite{Woos94}, but to the short timescale ($\tau \le 10$~ms) when
$L_\nu$ is still high and thus $\dot M$ is large, as can be seen in
Fig.~2. The third peak formation can be seen only for the proto-NS with
$M \ge 1.9 M_\odot$ \cite[with $R_\nu = 10$~km,][]{Wana01}, which (if
exist) might further collapse to a black hole. The ejecta mass of
$r$-processed material per event for model~C is estimated to be $\sim 1
\times 10^{-4} M_\odot$, which is in good agreement with the requisit
amount obtained from Galactic chemical evolution studies \cite{Ishi99,%
Ishi04}.

It should be noted that $R_\nu$ might be significantly larger than
$10$~km assumed here (for models~A and C), in particular at an early
phase with $L_\nu > {\rm several~} 10^{51}$~erg~s$^{-1}$ \cite[depending
on the equation of state for the nuclear matter (EOS)
applied,][]{Woos94, Prue05}, which would result in lower $S$ and longer
$\tau$. Hence the results shown in Fig.~3 do not guarantee the third
peak formation for $M = 2.0 M_\odot$ (nor the second peak formation for
$M = 1.4 M_\odot$). This should be taken as the {\it minimum} requisite
mass (or compactness $M/R_\nu$) of the proto-NS in order to obtain the
third $r$-process peak abundances.

\subsubsection{Neutrino Effects on Nucleosynthesis}

Possible effects of neutrino interactions on the $r$-process have been
extensively investigated by a number of authors \cite{Meye95, Full95,%
McLa96a, McLa96b, McLa97, Qian97, Meye98, Wana01, Kolb04, Tera04}. When
restricted to the physical conditions deduced from the ``realistic''
modelings of neutrino winds, however, the major contributors to the
$r$-process would be only the neutrino interactions on free nucleons and
on $\alpha$ particles. Other effects, i.e., the neutrino interactions on
heavy nuclei and subsequent neutron emission (or fission), are estimated
to be small because of their small cross sections
\cite[e.g.,][]{Tera04}, which might be buried with large uncertainties
in, e.g., nuclear physics far from $\beta$-stability as well as
astrophysical conditions \cite[e.g.,][]{Wana04}.

The neutrino capture on free nucleons affects the $r$-process by
changing $Y_e$ -- so called the ``$\alpha$ effect'' \cite{Full95,%
McLa96b, Meye98}. As the temperature decreases to $T_9 \sim 7$, $\alpha$
particles form by assembling from free neutrons and protons, while the
number ratio of free neutrons to free protons is locked by neutrino
capture on free nucleons in the intense neutrino flux. As a result, the
formation of $\alpha$ particles continues and $Y_e$ approaches $\sim
0.5$, which may hinder the $r$-process \cite{Meye98}. This plays,
however, only a minor (but non-negligible) role in the ``realistic''
neutrino winds. For example, the increase of $Y_e$ (= 0.40, initially)
from $T_9 = 9$ to 2.5 (at the onset of $r$-process) is no more than 0.03
for the winds in model~C (\S~2.1.1). Note that this effect is of
importance only at later phase ($L_\nu < {\rm several~}
10^{51}$~erg~s$^{-1}$), where the longer dynamic timescale as well as
the shorter distance from the neutrino sphere at $T_9 \sim 7$ results in
relatively larger neutrino fluence regardless of the lower $L_\nu$ (see
Fig.~2c).

Neutrino spallation reactions on $^4$He may also affect the
nucleosynthesis because of the large abundance of $\alpha$ particles in
the high-entropy wind \cite{Meye95}. This is due to the increase of seed
abundances even after the freezeout of three-body (i.e., $\alpha (\alpha
n, \gamma)^9{\rm Be}$ and $\alpha (\alpha \alpha, \gamma)^{12}{\rm C}$)
reactions at $T_9 \sim 3$, through the two-body reaction pathways opened
up by the spallations, e.g., $\alpha(\nu, \nu'p)^3$H followed by
$^3$H$(\alpha, \gamma)^7$Li and further $\alpha$ capture. As a result,
the $r$-process may be significantly hindered owing to the reduced
neutron-to-seed ratio, although its efficiency is highly dependent on
the neutrino spectra and luminosities as well as on the fluid dynamics
near the proto-neutron star \cite{Meye95, Tera04}. Note that neutrino
spallations of neutrons, $\alpha(\nu, \nu'n)^3$He, have no effect on the
nucleosyntheis in neutron-rich environment, which is immediately
followed by $^3$He$(n, \gamma)\alpha$ \cite{Meye95}.

\begin{figure}[t]
\begin{minipage}[t]{68mm}
\includegraphics*[width=\textwidth]{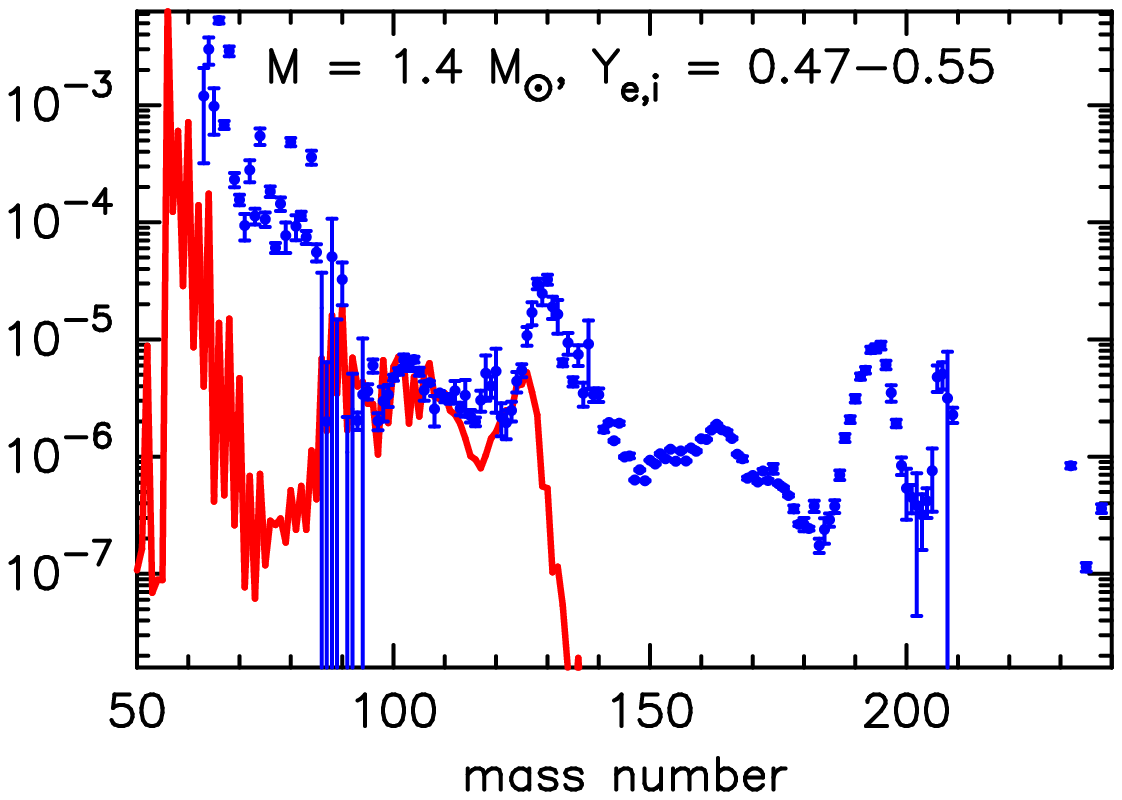} \caption{The final
abundances as a function of mass number averaged by the ejected mass and
$Y_e$ (see text) for $M = 1.4 M_\odot$. Also denoted are the scaled
solar $r$-process abundances (points).}
\end{minipage}
\hspace{\fill}
\begin{minipage}[t]{68mm}
\includegraphics*[width=\textwidth]{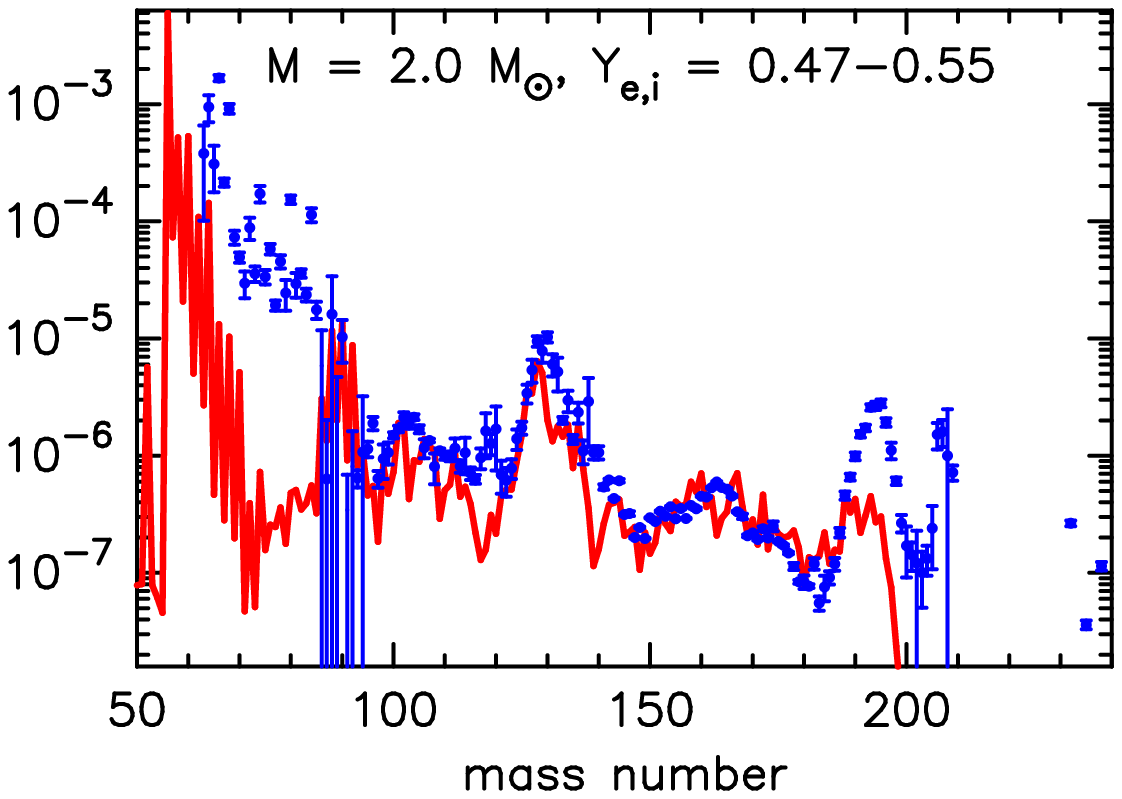} \caption{Same as Fig.~4, but
for $M = 2.0 M_\odot$.}
\end{minipage}
\end{figure}

\subsubsection{Overproduction Problem}

As can be seen in Figs.~3b-c and in many other ``successful''
$r$-process calculations, one worrisome aspect of the neutrino wind
scenario is a large overproduction of $N = 50$ (closed neutron shell, $A
\approx 90$) nuclei synthesized through $\alpha$-process by a factor of
$10-100$ \cite[e.g.,][]{Woos94, Wana01}. This originates from the
moderately high entropy ($50-100 N_A k$) ejecta with a large $\dot M$
before the $r$-process epoch ($L_\nu > 10^{52}$~erg~s$^{-1}$)
(Fig.~2). The overproduction diminishes when decreasing the neutron
richness in the wind to $Y_e \sim 0.49$ \cite{Hoff96, Frei99a,%
Wana02}. This is due to the termination of $\alpha$-process by
photodisintegration at $N \approx Z \approx 28$ rather than $N \approx
50$. Instead, some interesting isotopes $^{64}$Zn, $^{70}$Ge, and light
$p$-process nuclei $^{74}$Se, $^{78}$Kr, $^{84}$Sr, and $^{92}$Mo are
produced \cite{Hoff96, Wana05}, which seem difficult to be fulfilled by
other astrophysical sites \cite[but see possible explanations for
$^{64}$Zn,][]{Umed02, Prue05}. In fact, recent detailed hydrodynamic
simulations of ``successful'' SN explosions with accurate neutrino
transport show that $Y_e$ at early times is close to 0.5 or even higher
\cite{Prue05, Froh04, Bura05}. For example, a two-dimensional
hydrodynamic simulation by Buras et al. \cite{Bura05} shows that the
$Y_e$ values in the neutrino-processed ejecta during the early phase of
the explosion distribute between 0.47 and 0.56 with the maximum amount
at $\sim 0.50$.

As an excises, the nucleosynthesis results with the variation of $Y_e$
are presented, which was assumed to be a constant value (= 0.40) in
\S~2.1 \cite[for a more detailed discussion, see][]{Wana05}. Here, the
initial $Y_e$ is assumed to be constant ($Y_{e0}$) for $t_0 < t \le t_1$
and $Y_e (t) = (Y_{e0} - Y_{ef}) (t/t_1)^{-1} + Y_{ef}$ for $t > t_1$,
where $t_1 = 2$~s and $Y_{ef} = 0.35$, which mimics the hydrodynamic
results in Woosley et al. \cite{Woos94}. The values of $Y_{e0}$ are
taken to be from 0.47 to 0.55 (nine cases) according to Buras et
al. \cite{Bura05}, instead of $\sim 0.46$ in Woosley et
al. \cite{Woos94}. The mass-averaged nucleosynthesis results as in
\S~2.1 are further $Y_e$-averaged with the $Y_{e0}$ distribution of
neutrino-processed ejecta obtained by Buras et
al. \cite[][Fig.~38]{Bura05}. The final abundance curve is shown in
Figs.~4 and 5 as a function of mass number, which is compared to the
scaled solar $r$-process abundances.

As can be seen, no overproduction of the $A \approx 90$ nuclei appears
in this abundance curve for both $M = 1.4 M_\odot$ and $2.0 M_\odot$
cases. In fact, the harmful overproduction at $N = 50$ is now replaced
with the appropriate production of the $p$-nuclei $^{92}$Mo
\cite{Wana05}. This is due to the dominance of the matter with $Y_{e0}
\ge 0.49$ at the early phase of neutrino winds assumed here. However,
the amount of the $r$-processed material seen in Fig.~5 is about one
order smaller (and the third peak abundances are also deficient in this
case) than that in Figure~3c, since the $r$-process does not take place
when $L_\nu$ is high (and thus $\dot M$ is large) as in \S~2.1, owing to
the high $Y_e$ at the early phase. Given the neutrino wind is the major
production site of the $r$-process nuclei, therefore, it is not evident
if merely the proton richness in the neutrino-heated ejecta at the early
phase solves the overproduction problem.

\subsubsection{Is the Answer ``Blowing in the Wind''?}

A most probable implication is that the neutrino winds from a {\it
typical} proto-NS (e.g., $M = 1.4 M_\odot$ and $R_\nu = 10$~km) are
responsible for the production of {\it only} light $r$-process nuclei
such as Sr, Y, and Zr, and no heavier than the second peak ($A = 130$)
as can be seen in Fig.~3a, with some interesting isotopes (e.g.,
$^{64}$Zn, $^{70}$Ge, and light $p$-process nuclei) between $A = 60$ and
100 \cite{Wana05}. This is still of importance, however, since there are
increasing evidences that at least two different astrophysical sites
exist for the origins of ``light'' and ``heavy'' $r$-process nuclei (see
\S~3.3). Nevertheless, a possibility of the production of species beyond
the second ($A = 130$) and third ($A = 195$) peaks with a very {\it
compact} proto-NS (e.g, $M/R_\nu = 0.2 M_\odot/$km as for models~B and C
in Figs.~2 and 3) cannot be ruled out. In fact, many EOSs meet this
condition, $M/R_\nu \approx 0.2 M_\odot/$km, near their maximum masses
\cite[$M \approx 2.0 - 2.3 M_\odot$, see][]{Latt01, Wana01}. Recent
measurements of NS masses in binary systems also support the presence of
such massive NSs \cite{Latt04}. It should be noted that a proto-NS's
mass could be slightly larger than the maximum mass of a cold star
because of its extra leptons and thermal energy. In this case, collapse
to a black hole would take place (after the $r$-process) on a diffusion
time of a few 10~s, which might have occured in SN~1987A \cite{Latt04}.

If the neutrino winds were {\it really} the major production site of the
heavy $r$-process nuclei, therefore, the progenitor would have a
relatively large mass, e.g., $\ge 20 M_\odot$. On the other hand, the
lighter $r$-process nuclei would be supplied from low mass progenitors
($\sim 10 - 15 M_\odot$). This difference may reflect the change of a
core structure with the progenitor mass, i.e., the steep density
gradient with the small iron core ($\sim 1.3 M_\odot$) for a star of
$\le 15 M_\odot$ and the mild density gradient with the massive iron
core ($\sim 1.8 M_\odot$) for a star of $\ge 20 M_\odot$
\cite{Burr95}. It should be noted that the very massive progenitors
would suffer from a significant ``fallback'' of the matter once ejected,
resulting in, perhaps, no ejection of $r$-processed material
\cite{Thom01}. Hence, the progenitors for the origin of the heavy
$r$-process nuclei may be limited to a small mass range, e.g., $20 - 25
M_\odot$. This would make such an event relatively rare, accounting only
about $10 \%$ of all core-collapse SN events. This (moderate) rareness
does not cause a problem, but rather is needed from Galactic chemical
evolution as discussed in \S~3.2.

It should be emphasized that the implications above are all based on the
assumption of spherical symmetry as well as on the arbitrary chosen
$Y_e$. There have been no qualitative studies of $r$-process in
asymmetric neutrino winds nor with an accurate determination of
$Y_e$. Therefore, conclusions described here might be modified by the
future works based on more {\it realistic} modelings of neutrino winds
with multidimensional hydrodynamics as well as with an accurate
treatment of neutrino transport. It is interesting to note that recent
two-dimensional simulations demonstrate that hydrodynamic instabilities
can lead to low-mode ($l = 1,\, 2$) asymmetries of the fluid flow in the
neutrino-heated layer behind the SN shock \cite{Sche04, Jank05}. This
provides not only a natural explanation for aspherical mass ejection and
for pulsar kicks but shows some promise as the yet unknown explosion
mechanism of core-collapse SNe \cite{Jank05}. Such multi-dimensional
effects may have to be taken into account in the future work, since the
$r$-process takes place relatively close to the core ($\sim
100-1000$~km) where the asymmetry plays a significant role. The strong
magnetic field (``magneter-like'' strength such as $\sim 10^{15}$~G,
three orders of magnitude larger than the typical value) in a proto-NS
has been also suggested to increase entropy and thus help the
$r$-process even with $M \approx 1.4 M_\odot$ \cite{Thom03, Suzu05}
\cite[but see][]{Ito05}. Such SN events account for no more than a few
\% of all SN events. This might be, however, still in reasonable
agreement with the constraint from Galactic chemical evolution (\S~3.2).

\subsection{Prompt Explosion}

If a massive star explodes hydrodynamically at core bounce prior to the
delayed neutrino heating, the ejecta keeps its neutron richness due to
electron capture, $Y_e \sim 0.2$, which may lead naturally to
$r$-processing regardless of the relatively low entropy, $S \sim 10 N_A
k$ \cite{Schr73, Sato74, Hill76}. This is one of the reasons that this
scenario, ``prompt explosion'' has been still considered to be a
possible explanation for the $r$-process origin \cite{Whee98, Sumi01,%
Wana03}, despite difficulties in achieving such an explosion by
self-consistent hydrodynamic calculations. In fact, many previous works
have suggested that even the SNe near their lower-mass end ($\sim 10
M_\odot$), which form small iron cores ($\sim 1.3 M_\odot$), have
difficulties in achieving hydrodynamic explosions \cite{Bowe82, Burr83,%
Burr85, Brue89a, Brue89b, Baro90}. Optimistically saying, a prompt
explosion may occur in the collapse of the {\it lowest} mass
progenitor, perhaps an $8-10 M_\odot$ star that forms an O-Ne-Mg core at
its center, owing to its small gravitational potential as well as the
small NSE core at the onset of core bounce \cite{Nomo84, Hill84,%
Miya87}.

\begin{figure}[t]
\begin{minipage}[t]{68mm}
\includegraphics*[width=\textwidth]{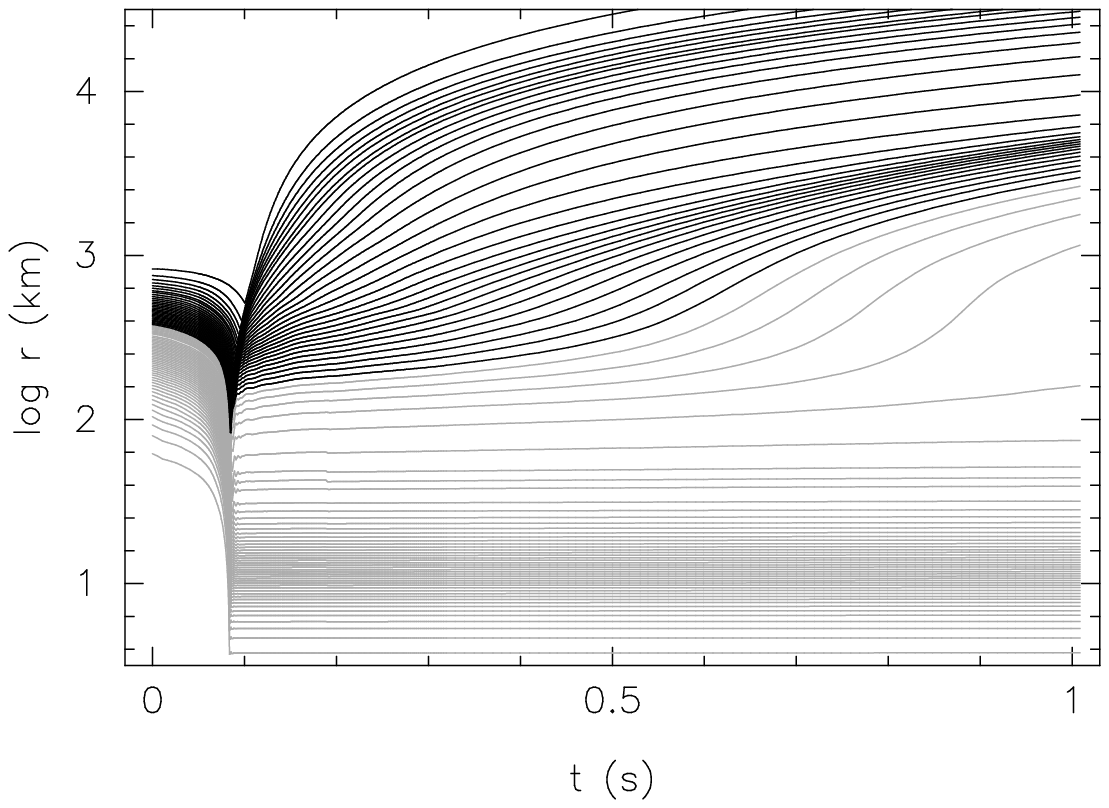} \caption{Time variations of
radius for selected mass points in the prompt explosion of a $9 M_\odot$
star, in which the shock-heating energy is enhanced {\it artificially}
by a factor of 1.6.}
\end{minipage}
\hspace{\fill}
\begin{minipage}[t]{68mm}
\includegraphics*[width=\textwidth]{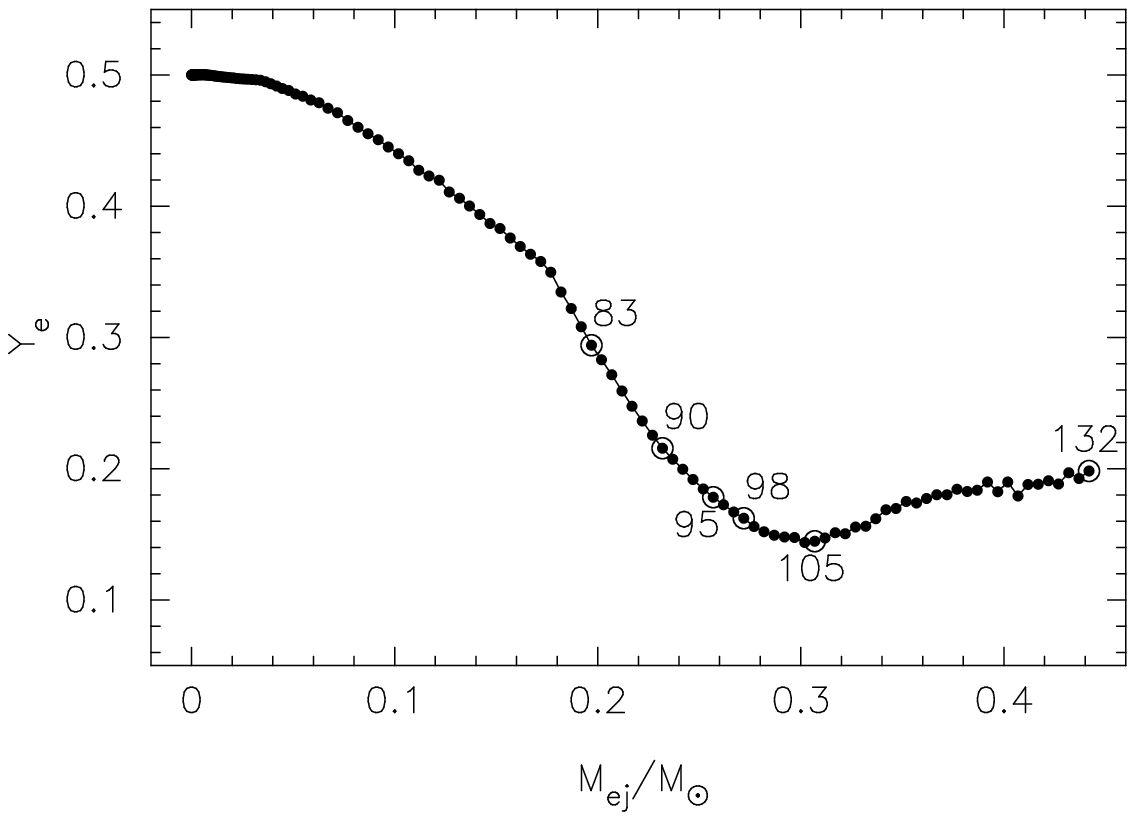} \caption{$Y_e$ distribution
in the ejected material. The surface of the O-Ne-Mg core is at mass
coordinate zero. Some selected mass points are denoted by zone numbers
(see Fig.~8).}
\end{minipage}
\end{figure}

It should be noted that recent detailed core-collapse simulations of the
$9 M_\odot$ star having an O-Ne-Mg core with accurate treatment of
neutrino transport \cite{Lieb04, Jank05} do not confirm the prompt
explosion found in a previous study with simpler neutrino treatment
\cite{Hill84}. Instead, a weak explosion by delayed neutrino-heating
emerges \cite[this is only one case that a one-dimensional
self-consistent simulation with accurate neutrino transport results in
an explosion,][]{Jank05}. In addition, the fate of the stars in this
mass range is quite uncertain, which is highly dependent on the
treatment of the convection as well as the mass loss assumed in the
calculations of stellar evolution. In particular, an efficient mass loss
would result in losing whole the envelope before reaching the
Chandrasekhar mass and then leaving an O-Ne-Mg white dwarf
\cite{Nomo84}. As a result, the mass range of the stars that undergo
SNe would be restricted between $M_{\rm WD}$ and $10 M_\odot$, where $8
M_\odot \le M_{\rm WD} \le 10 M_\odot$ \cite[see also recent
studies][]{Rito96, Iben97, Rito99, Eldr04}. A limited mass range, say,
between $M_{\rm WD} = 9.5 M_\odot$ and $10 M_\odot$, still accounts for
about $7-8 \%$ of all core-collapse SN events, which is in good
agreement with a constraint from Galactic chemical evolution
(\S~3.2). Hence, comprehensive studies including stellar evolutions
covering whole this mass range, as well as the subsequent core-collapse
simulations, are awaited before drawing any final conclusions. In the
meantime, however, it would be valuable to examine the $r$-process
nucleosynthesis in a {\it schematic} prompt explosion forced by, e.g.,
enhancing shock-heating energy \cite{Wana03} or suppressing electron
capture \cite{Sumi01}. In the following, our recent result on the
$r$-process in a collapsing O-Ne-Mg core \cite{Wana03} is briefly
presented \cite[see also][]{Sumi01}.

A {\it purely} hydrodynamical (i.e., without neutrino) core-collapse
simulation of a $9M_\odot$ star \cite{Nomo84} that forms a 1.38
$M_\odot$ O-Ne-Mg core is performed with a one-dimensional implicit
Lagrangian hydrodynamic code with Newtonian gravity. Major input physics
is the equations of state of nuclear matter (EOS) \cite{Shen98} and of
the electron (and positron) gas with arbitrary relativistic pairs as
well as arbitrary degeneracy, and electron (and positron) capture on
free nucleons and nuclei \cite{Lang00}. The capture is suppressed above
$\rho = 3 \times 10^{11}$g~cm$^{-3}$ to mimic the {\it neutrino
trapping}. The composition of the O-Ne-Mg core is held fixed until the
temperature reaches $T_9 = 2$ that is taken to be the onset of oxygen
burning, at which point the matter is assumed to instantaneously be in
NSE. We find that only a weak explosion results with the explosion
energy of $E_{\exp} = 1.8 \times 10^{49}$~ergs, where the minimum $Y_e$
is only 0.45 and no $r$-processing is expected. In order to examine the
possible operation of the $r$-process in this star, an energetic
explosion ($E_{\exp} = 3.5 \times 10^{51}$~ergs) is {\it artificially}
obtained by multiplying a factor of 1.6 to the shock-heating term in the
energy equation (Fig.~6). The highly neutron-rich matter ($Y_e \approx
0.14$, Fig.~7) from deeper inside of the core is ejected, which results
in robust $r$-processing as can be seen below.

\begin{figure}[t]
\begin{center}
\includegraphics*[width=\textwidth]{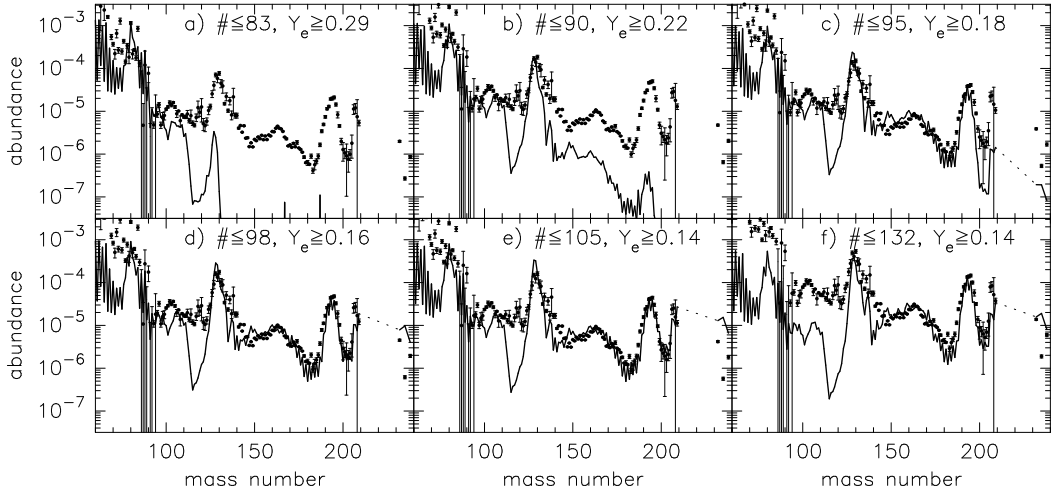}
\end{center}
\caption{Mass-averaged $r$-process abundances (line) as a function of
mass number obtained from the ejected zones in (a) $\le 83$, (b) $\le
90$, (c) $\le 95$, (d) $\le 98$, (e) $\le 105$, and (f) $\le 132$ (see
Fig.~7). These are compared with the solar $r$-process abundances
(points) \cite{Kapp89}, which is scaled to match the height of the first
peak ($A = 80$) for (a), the second peak ($A = 130$) for (b), and the
third peak ($A = 195$) for (c)-(f).}
\end{figure}

The yields of $r$-process nuclei obtained with the nuclear reaction
network (Fig.~1) are mass-averaged from the surface (zone 1) to the
zones (a) 83, (b) 90, (c) 95, (d) 98, (e) 105, and (f) 132 (see Fig.~7),
which are compared with the solar $r$-process abundances \cite{Kapp89}
as can be seen in Fig.~8. A solar $r$-process pattern for $A \ge 130$ is
naturally reproduced in cases~c-f, owing to the ejection of highly
neutron-rich matter ($Y_e < 0.20$). On the other hand, the solar-like
abundance curves up to $A \approx 100$ and 130 in cases~a and b,
respectively, can be seen. Note that a problematic overproduction of $N
= 50$ ($A \approx 90$) nuclei (\S~2.1.3) cannot be seen for all the
cases (Figs.~8a-f). This is due to the low entropy ($\sim 10 N_A k$) of
the shock-heated matter (without neutrino-heating), in which the
$\alpha$-rich freezeout (that can be seen in neutrino winds) is not of
significance. The deficiency of $r$-process nuclei at $A \sim 115$
reflects the strong shell gap at $N = 50$ in the nuclear mass formula
adopted to deduce the neutron-capture rates in this mass range. This
valley might be fulfilled with another nuclear mass formula (see
\S~2.4) or another astrophysical site (e.g., neutrino winds, see
Fig.~3c). Given the ejecta mass $M_{\rm ej}$ is reduced because of,
e.g., a weaker explosion or fallback of the once ejected matter by the
reverse shock, the prompt explosion from a collapsing O-Ne-Mg core
considered here can be regarded as the origin of either ``light''
(cases~a-b), ``heavy'' (case~f) , or ``all'' (cases~c-e) $r$-process
nuclei. It is interesting to note that the production of thorium and
uranium differs from model to model, even though the abundance pattern
seems to be {\em universal} between the second and third $r$-process
peaks. It should be noted that this event may not be the origin of
nuclei lighter than $A \sim 70$. The mass of ejected iron is only
$\approx 0.02 M_\odot$, and the production of $\alpha$ nuclei is
negligible, since the outer envelope consists of, if survived from mass
loss, only hydrogen and helium layers.

A serious problem in this scenario, other than {\it if it explodes}, is
the {\it overproduction} of the ``total'' $r$-processed matter. While
the abundance distribution is in good agreement with the solar
$r$-process curve without an overproduction of $A \approx 90$ nuclei as
seen in the neutrino wind, the ejected mass of $r$-processed matter in,
e.g., case~e in Fig.~8 is about $0.05 M_\odot$. This is more than two
orders of magnitude larger than the requirement from Galactic chemical
evolution (a few $10^{-4} M_\odot$). In addition, the remnant mass in
this case results in only $1.13 M_\odot$ that is significantly smaller
than the ``typical'' mass $1.4 M_\odot$, although a few NSs with
measured masses are suspected to have such low masses \cite[with
relatively large errors, see][]{Latt04}. A possible explanation for this
problem is that only a small fraction ($\sim 1 \%$) of $r$-process
material is ejected by ``mixing-fallback'' of the core matter
\cite{Umed02, Umed03}, wherein most of the $r$-process material falls
back onto the proto-neutron star. An asymmetric explosion, such as that
from rotating cores or jets may have a similar effect as the ejection of
deep-interior material in a small amount. If this happened, the typical
mass of the proto-NS ($\approx 1.4 M_\odot$) would be recovered.

\subsection{Other Scenarios}

The appearance of $r$-process elements in the old halo stars in the
Galaxy no doubt demands the $r$-process nuclei to have a primary origin
(\S~3), wherein the seed nuclei for neutron capture can be synthesized
by itself as in the neutrino wind and the prompt explosion. In this
regard, additional astrophysical sites that show some promise as the
$r$-process origin currently suggested are the ``NS mergers'', the
``accretion-induced collapses (AIC)'', and the ``collapsars''.

Of particular importance among these scenarios would be the coalescence
of two NSs (or of an NS and a black hole), i.e., the ``NS merger'',
which might naturally provide the neutron-rich environment needed for
$r$-process. The presence of double NS binaries with extremely short
periods \cite[e.g.,][]{Latt04} is an indirect, but unambiguous evidence
that such events exist in reality, although its event rate is poorly
known. So far, little effort has been devoted to the nucleosynthetic
study in this event \cite{Meye89, Frei99b, Gori05a}, which would be
premature to make any firm predictions of its contribution to Galactic
chemical evolution. Nevertheless, recent studies suggest a solar-like
$r$-process abundance production for nuclei with $A > 130$ in such
events \cite{Frei99b, Gori05a} and no lighter nuclei than $A \sim 70$
\cite{Gori05a}, which might be distinguishable from abundance
determinations of extremely metal-poor halo stars.

An AIC of a (C-O or O-Ne-Mg) white dwarf in a close binary system
\cite{Nomo91} is an analogous event to a core-collapse SN, resulting in,
perhaps, similar outcome to that of the neutrino wind or the prompt
explosion. A lack of the outer envelope may result in, however, the
production of no $\alpha$ and little iron-peak elements similar to the
prompt explosion \cite{Qian03, Wana03}. Note that the presence of a
dense accretion disk around the core may help the matter to be
neutron-rich even in the neutrino-heated ejecta. There has been,
however, no $r$-process abundance prediction so far, and a quantitative
study in the future is highly desired. A collapsar is also suspected to
be an astrophysical $r$-process site, owing to its extremely high entropy
along the polar direction as well as the dense accretion disk with low
$Y_e$ around the nascent black hole \cite{MacF99, Prue04}. Since the
central engine that drives the jets to induce a gamma-ray burst or a
hypernova is still unknown, it would be too early to state any
predictions here, and future quantitative studies are also desired.

\subsection{Uncertainties in the nuclear data far from $\beta$-stability}

Besides astrophysical conditions described above, another underlying
difficulty for $r$-process calculations is due to the uncertainties in
the theoretical predictions of nuclear data far from the
$\beta$-stability, for which essentially no experimental data exist
\cite[for a recent review, see][]{Lunn03}. In particular, mass
predictions for neutron-rich nuclei play a key role since they affect
all the nuclear quantities of relevance in the $r$-process, namely the
neutron capture, photodisintegration and $\beta$-decay rates, as well as
the fission probabilities.

\begin{figure}[t]
\begin{center}
\includegraphics*[width=\textwidth]{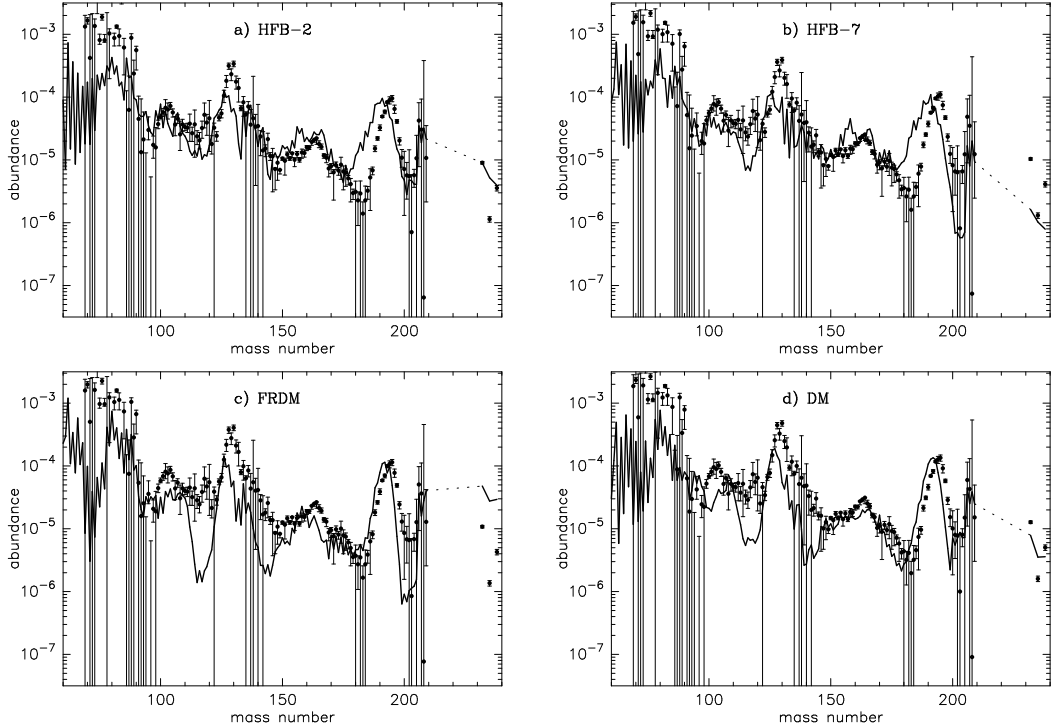}
\end{center}
\caption{Final mass-averaged $r$-process abundances (line) as a function
of mass number obtained with various mass formulae; (a) HFB-2, (b)
HFB-7, (c) FRDM, and (d) DM.  These are compared with the solar
$r$-process abundances (points), which are scaled to match the height of
the third $r$-process peak.  }
\end{figure}

Attempts to estimate nuclear masses go back to the liquid-drop
Weizs\"acker mass formula. Improvements to this original model have been
brought little by little, leading to the development of
macroscopic-microscopic mass formulae, such as the droplet model
\cite[e.g.,][DM]{Hilf76} and the ``finite-range droplet model'' (FRDM)
of \cite{Moel95}. In this framework, the macroscopic contribution to
the masses and the microscopic corrections of phenomenological nature
are treated independently, both parts being connected solely through a
parameter fit to experimental masses. As a consequence, its reliability
when extrapolating far from experimentally known masses is severely
limited, despite the great empirical success of these formulae
(e.g. FRDM fits the 2135 $Z\ge 8$ experimental masses \cite{Audi95} with
an rms error of 0.676 MeV). A new major progress has been achieved
recently within the Hartree-Fock-Bogoliubov (HFB) method \cite{Samy03,
Gori02, Gori03, Gori05b}. It is now demonstrated that this fully
microscopic approach, making use of a Skyrme force fitted to essentially
all the experimentally known mass data, is not only feasible, but can
successfully compete with the most accurate droplet-like formulae
available nowadays (e.g., FRDM) in the reproduction of measured masses
(e.g., an rms error of the order of 0.674~MeV for the HFB-2
\cite{Gori02} mass table).

Most particularly, the HFB mass formulae show a weaker neutron-shell
closure close to the neutron drip line with respect to droplet-like
models such as FRDM. This effect can be seen in Fig.~9 that shows the
results of $r$-process calculations with four sets of nuclear mass
formulae HFB-2 \cite{Gori02}, HFB-7 \cite{Gori05a}, FRDM \cite{Moel95},
and DM \cite{Hilf76}, with the prompt explosion model described in
\S~2.2 \cite[for more detail, see][]{Wana04}. Due to its weak shell
effect at the neutron magic numbers in the neutron-rich region, the
microscopic mass formulae (HFB-2 and HFB-7) give rise to a spread of the
abundance distribution in the vicinity of the $r$-process peaks ($A =
130$ and 195). While this effect resolves the large underproduction at
$A \approx 115$ and 140 obtained with droplet-type mass formulae (FRDM
and DM), large deviations compared to the solar pattern are found near
the third $r$-process peak. When using the droplet mass formulae, sharp
$r$-process peaks are systematically found, owing to their strong shell
effect for neutron magic numbers even in the neutron-rich
region. However, due to the numerous uncertainties still affecting the
astrophysics models as well as the prediction of extra nuclear
ingredients, it would be highly premature to judge the quality of the
mass formula on the basis of such a comparison.

\begin{figure}[t]
\begin{center}
\includegraphics*[width=\textwidth]{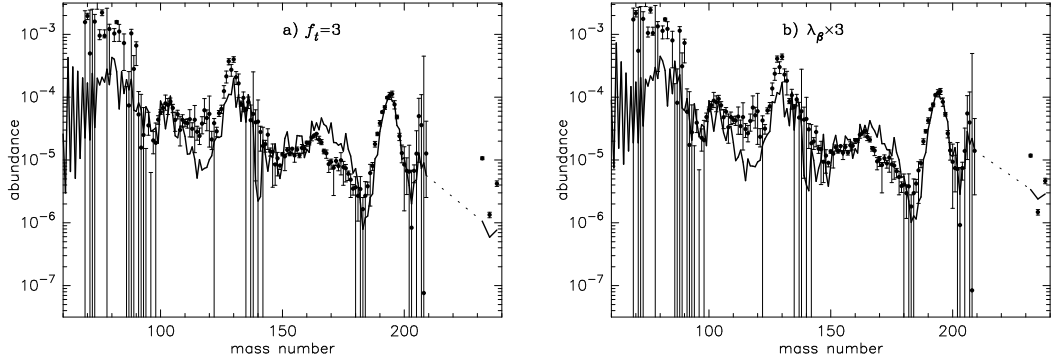}
\end{center}
\caption{ Same as Fig.~9, but for (a) slow trajectories and (b) fast
$\beta$-decay rates (a factor of three, see text) with the HFB-2 mass
formula.  }
\end{figure}

For example, we find that abundance peaks similar to the one observed in
the solar system could be recovered with the HFB-2 mass formula if the
dynamical timescales of the mass trajectories are increased by a factor
of three (without any change in the entropy, Fig.~10a) or by decreasing
systematically the $\beta$-decay half-lives by the same factor
(Fig.~10b). This is a consequence that the freezeout of $(n,\, \gamma)$
and $(\gamma,\, n)$ reactions takes place at higher temperature (and
thus closer to the $\beta$-stability), where the shell gaps at neutron
magic numbers are evident in the HFB-2 masses \cite[for more detail,
see][]{Wana04}. These changes might be conceivable when considering the
current uncertainties in the astrophysics as well as in the nuclear
$\beta$-decay model. Much effort in the astrophysics and nuclear
modeling remain to be devoted to improve the difficult description of
the $r$-process nucleosynthesis.

\section{Galactic Chemical Evolution}

While no consensus has been achieved on the astrophysical $r$-process
site from the nucleosynthetic point of view, Galactic chemical evolution
studies provide several important clues to this puzzle from another
point of view, when combined with recent comprehensive spectroscopic
analyses of extremely metal-poor halo stars. Major issues here are
threefold. First is the {\it universality} of the stellar abundance
distributions that agree with the scaled solar $r$-process curve at
least between the second and third $r$-process peaks ($Z \approx 56 -
78$). This implies uniqueness of the physical conditions to some extent,
in which the $r$-process proceeds. Second is huge star-to-star {\it
dispersion} of the $r$-process abundances relative to iron, which may
pose a significant constraint on the stellar mass range of the SN
progenitors as the origin of $r$-process nuclei. Third is the
disagreement of the lighter ($Z < 56$) neutron-capture elements with the
scaled solar $r$-process curve that match the heavier, which implies the
presence of at least {\it two} $r$-process sites.

\subsection{``Universality'' of the $r$-Process Abundances}

One of the most remarkable findings related to the spectroscopic studies
of Galactic halo stars in the last decade is the discovery of several
extremely metal-poor ([Fe/H]\footnote{$[A/B] \equiv \log (N_A/N_B) -
\log (N_A/N_B)_\odot$, where $N_A$ indicates abundance of $A$.} $\sim -
3$), $r$-process-enhanced ([Eu/Fe]\footnote{Eu is often taken to be
representative of $r$-process elements, since $94\%$ of its solar
abundance originates from $r$-process \cite{Arla99}.} $\sim 1-2$) stars,
whose abundances of neutron-capture elements are in excellent agreement
with the scaled solar $r$-process curve \cite{Sned96, Cayr01, Hill02,%
Cowa02, Sned03, Hond04, Chri04}. As can be seen in Fig.~11, the
neutron-capture element abundances in CS~22892-052 ([Fe/H] $= - 3.1$ and
[Eu/Fe] $= 1.7$) \cite{Sned96, Sned03} show an outstanding concordance
with the scaled solar $r$-process abundance curve, in particular between
the second and third $r$-process peaks ($Z = 56-82$). The appearance of
{\it purely} $r$-processed matter\footnote{$s$-process-enhanced,
extremely metal-poor {\it carbon} stars are not considered here, whose
atmosphere might have been polluted from the former AGB companions in
binaries \cite{Wana06}.} in the atmosphere of such old halo stars in
the Galaxy strongly support the idea that the production of $r$-process
nuclei is associated to short-lived massive stars, perhaps,
core-collapse SNe \cite{Trur81}. Furthermore, the uniqueness of the
abundance patterns of neutron-capture elements demonstrates the {\it
universality} of the $r$-process nucleosynthesis that occurs in,
perhaps, a unique astrophysical site.

\begin{figure}[t]
\begin{minipage}[t]{68mm}
\includegraphics*[width=\textwidth]{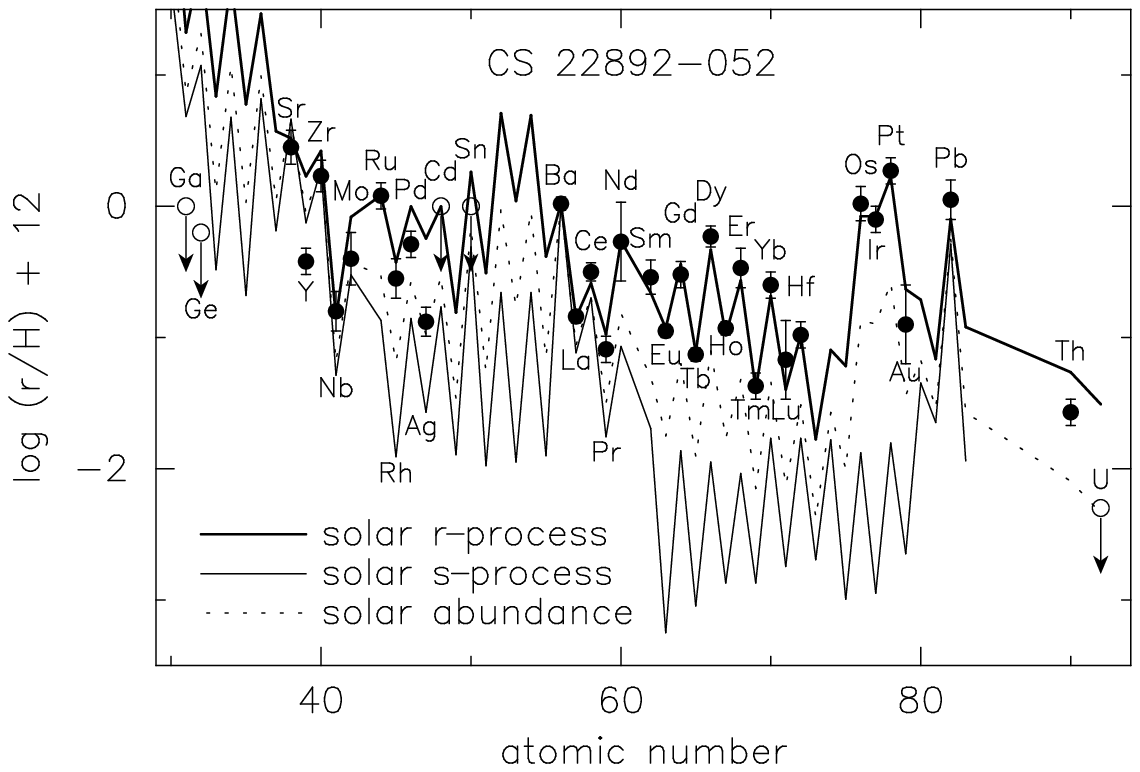} \caption{Observed abundances
in CS~22892-052 \cite{Sned03}. The metallicity of this star is [Fe/H] $=
- 3.1$. Detected elements are shown as filled circles with error bars,
and upper limits are denoted with open circles. The solar $r$-process
(thick-solid line), $s$-process (thin-solid line), and the solar
abundances (dotted line) \cite{Ande89} are vertically scaled to match
the observed Ba abundance.}
\end{minipage}
\hspace{\fill}
\begin{minipage}[t]{68mm}
\includegraphics*[width=\textwidth]{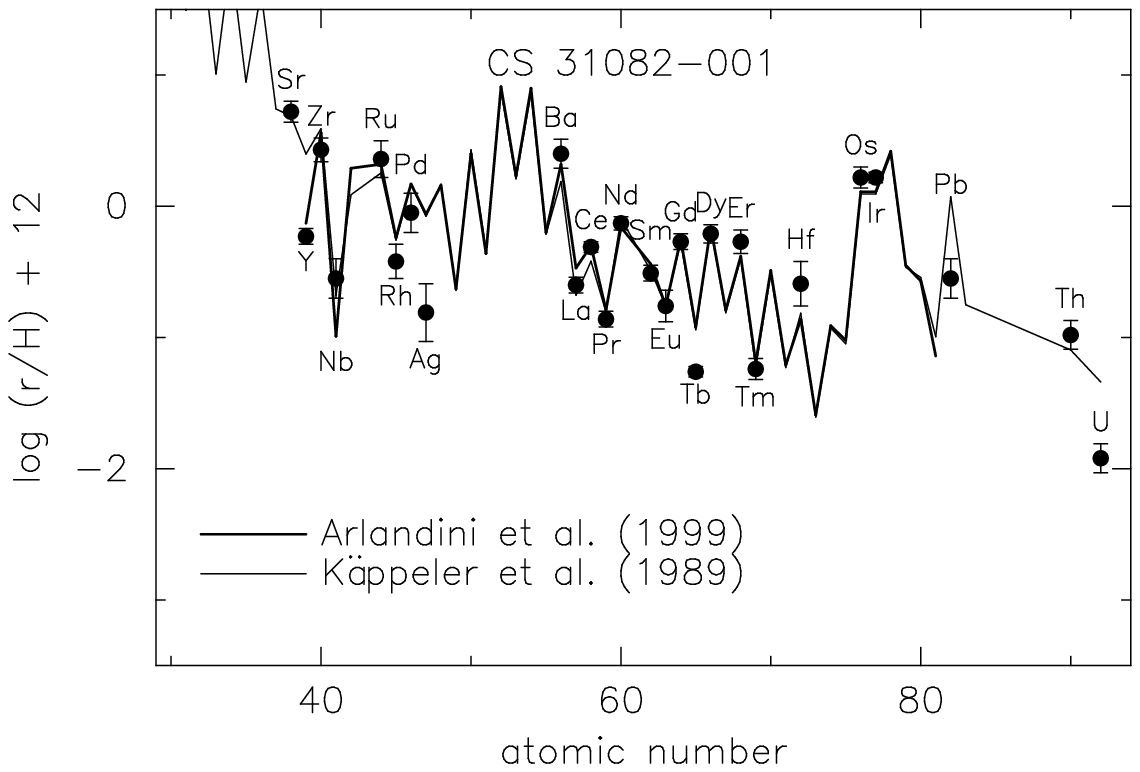} \caption{Observed abundances
in CS~31082-001 \cite{Hill02, Ivar03, Plez04}. The metallicity of this
star is [Fe/H] $= - 2.9$. Detected elements are shown as filled circles
with error bars. The solar $r$-process abundances from Arlandini et al.
\cite[][thick line]{Arla99} and K\"appeler et al. \cite[][thin
line]{Kapp89} are vertically scaled to match the observed Eu abundance.}
\end{minipage}
\end{figure}

Another notable discovery is the detection of uranium in CS~31082-001
([Fe/H] $= - 2.9$ and [Eu/Fe] $= 1.6$) \cite{Cayr01, Hill02}, which can
be regarded as a precise cosmochronometer in addition to Th previously
used for age dating \cite{Cowa97, Cowa99, Gori99} (\S~3.4). On the other
hand, however, the significantly high Th and U (and low Pb) abundances
compared to those in CS~22892-052 clearly show that the universality
does {\it not} hold beyond the third $r$-process (Pt) peak, which makes
the age dating assuming the ``universality'' of the $r$-process
abundance pattern questionable. This {\it non-universality} of the
$r$-process beyond the Pt-peak has been further confirmed by additional
findings of Th-rich stars \cite{Hond04, Yush05}.

It is not clear from currently available data that the {\it
universality} holds down to the elements near the first $r$-process
peak, e.g., Sr, Y, and Zr, owing to the deficiencies of a few elements
between the first and second $r$-process peaks, as can be seen in
Figs.~11 and 12. It should be noted that the $s$-process dominated
elements in the solar system, e.g., Sr (85\%), Y (92\%), and Ba (81\%),
involve large uncertainties when deriving the $r$-process components
from the observed solar values by subtracting the theoretically
calculated $s$-process contribution \cite{Gori97}. In fact, the
deficiency of Y abundance is cured when adopting the recent data from
Arlandini et al. \cite{Arla99} instead of the older table from
K\"appeler et al. \cite{Kapp89}, as can be seen in Fig.~12. The
deficiency of Ag relative to the scaled solar $r$-process curve can be
seen in {\it all} metal-poor stars that have its measured values
\cite{John02b, Ishi05}. Hence the low Ag abundance does not necessarily
indicate the break down of the universality below the second $r$-process
peak as previously suggested \cite[e.g.,][]{Sned03}, although the reason
of its deficiency is unknown\footnote{The uncertainty in deriving the
solar $r$-process component of Ag abundance may be small \cite{Gori97},
owing to its dominance (80\%) in the solar system abundance.}.

\subsection{``Dispersion'' of the $r$-Process Abundances}

Another striking feature of the observed neutron-capture element
abundances in extremely metal-poor stars is their large dispersions that
cannot be seen in any other elements \cite{Wool95, McWi95, Ryan96}. In
Fig.~13, the observed Eu abundances (as representative of heavy
$r$-process elements) relative to iron taken from the recent literature
are plotted. The dispersion for the measured values ranges about two
orders of magnitude at [Fe/H] $\sim - 3$, which is in contrast to the
exceedingly small scatters for $\alpha$ and iron-peak elements
\cite{Cayr04}.

This large dispersion may be interpreted as a result of incomplete
mixing of the interstellar medium (ISM) at the beginning of the
Galaxy. In the {\it standard} chemical evolution models that are
commonly used \cite[e.g.,][]{Timm95}, observed stellar compositions are
taken to represent those of the ISM averaged over whole the Galaxy when
the stars were formed. It may not be true, however, if star formations
are affected by nearby SNe. The composition of the newly formed star must
be a mixture of the low-metallicity ISM and the {\it single} (or a few)
SN ejecta with the high metal content. In the following, our recent
results of Galactic chemical evolution studies of $r$-process elements
are presented, along with our recent spectroscopic analysis of several
extremely metal-poor stars using SUBARU/HDS \cite[see][for more
detail]{Ishi99, Ishi04, Ishi05}. A few other recent studies that have
taken the effect of inhomogeneity in ISM into account show qualitatively
similar results \cite{Trav99, Tsuj00, Fiel02, Arga04}.

In our study, the evolutions of the ISM in the Galactic halo are
calculated by a one-zone (i.e., {\it homogeneous}) model as in the {\it
standard} approach, which loses gas through accretion onto the disk
\cite{Ishi99}. The star formation and accretion rates are assumed to be
proportional to the gas fraction of the halo. The star formations obey
the Salpeter initial mass function in the mass range $0.05-60
M_\odot$. The coefficients for the accretion rate and the star formation
rate are adjusted to fit to the observational data of [O/Fe] versus
[Fe/H] \cite[e.g.,][]{Barb88, Edva93} and the metallicity distribution
of halo stars \cite{Sand87}.

The chemical compositions of newly formed stars are determined as
follows. We assume that star formation is initiated by SNe. An SN
remnant is supposed to expand spherically until reaching the merge
radius with the ISM \cite[typically $\sim 100$~pc;][]{Ciof88}. At this
point, about $\sim 10^4 M_\odot$ of the ISM is swept up by the SN
remnant. The composition of a formed star is then assumed to be the mass
average of this ``snowplowed'' ISM and the {\it single} SN ejecta. The
mass of the SN progenitor is chosen randomly but obeying the initial
mass function. Contribution of this individual SN yield to the ISM is
not considered here, whose evolution is calculated independently by the
one-zone model.

\begin{figure}[t]
\begin{center}
\includegraphics*[width=\textwidth]{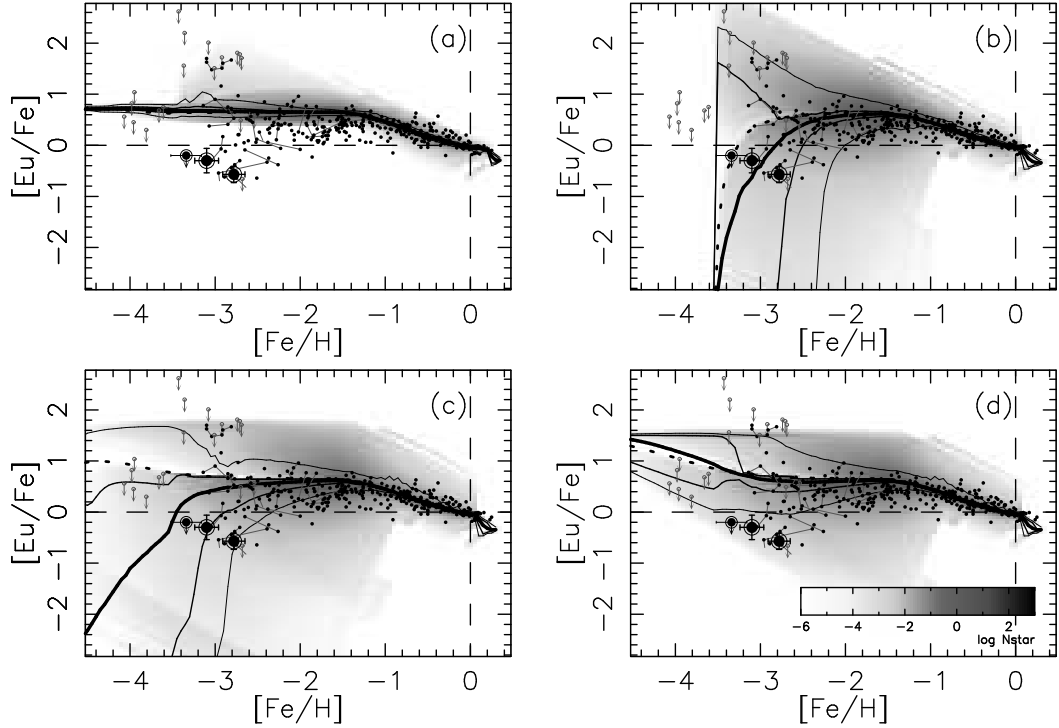}
\end{center}
\caption{Model predictions of [Eu/Fe] in stars as functions of [Fe/H]
are compared with the recent observations. The $r$-process site is
assumed to be SNe of (a) $\ge 10 M_\odot$, (b) $8-10 M_\odot$, (c)
$20-25 M_\odot$, and (d) $\ge 30M_\odot$. The predicted number density
of stars per unit area is shown by the grey images. The average stellar
abundance distributions are indicated by thick-solid lines with the 50\%
and 90\% confidence intervals (solid and thin-solid lines,
respectively). The average abundances of the ISM are denoted by the
thick-dotted lines. The observed abundances taken from the recent
literature \cite[filled and open circles for measured values and
upperlimits, respectively,][]{Grat94, McWi95, McWi98, Wool95, Ryan96,
Shet96, Sned96, West00, Burr00, Fulb00, Norr01, John02a, John02b,
Fran03, Hond04} are plotted, along with our recent data (large double
circles) obtained with SUBARU/HDS \cite{Ishi04}.}
\end{figure}

The iron yields for Type~II and Type~Ia SNe are taken from
\cite{Nomo97a} and \cite{Nomo97b}, respectively. The possible
metallicity effects are not considered here for simplicity\footnote{A
large metallicity dependence of the iron yield would result in too large
scatters of $\alpha$ elements relative to iron \cite{Ishi03}, which
conflicts with the recent spectroscopic studies of metal-poor stars
\cite[e.g.,][]{Cayr04}.}. The production of iron in $8-10 M_\odot$
stars, which is estimated to be small by nucleosynthesis calculations
\cite{Wana03}, is neglected here. The $r$-process site is currently
unknown, but {\it assumed} here to be the core-collapse SNe (either
``neutrino winds'' or ``prompt explosions'') from the stars of (a) $\ge
10 M_\odot$ (all SNe), (b) $8-10 M_\odot$ (low-mass end of SNe), (c)
$20-25 M_\odot$ (intermediate-mass SNe), and (d) $\ge 30$ (high-mass end
of SNe)\footnote{These mass ranges are chosen as the representative four
cases. A small shift of the range (e.g., $10-11 M_\odot$ for case~b or
$25-30 M_\odot$ for case~c) would not change significantly the current
results.}. Each case accounts for (a) 100\%, (b) 28\%, (c) 11\%, and (d)
15 \% of all SN events. The Eu yield is taken to be constant\footnote{In
reality, the $r$-process yields must be dependent on the SN progenitor
masses. A mild dependence over all the SN mass range would, however,
result in only a small star-to-star scatter similar to Fig.~13a. A
strong progenitor mass dependence of the Eu yield would have a similar
effect to the restricted mass range with the constant Eu yield
considered here \cite[see, e.g.,][]{Tsuj00}.} for simplicity, over the
stellar mass range for each, (a) $1.2 \times 10^{-7} M_\odot$, (b) $3.1
\times 10^{-7} M_\odot$, (c) $1.1 \times 10^{-6} M_\odot$, and (d) $7.8
\times 10^{-7} M_\odot$, which is scaled to reproduce the solar value
[Eu/Fe] = [Fe/H] = 0, and to be zero outside of the range. It can be
seen that the limited mass range demands the high $r$-process material
ejected per event. The corresponding {\it total} mass of $r$-process
nuclei for each case is about a few $10^{-5} M_\odot$ to a few $10^{-4}
M_\odot$, which is in good agreement with theoretical estimates from
nucleosynthesis calculations in neutrino winds \cite{Woos94, Wana01,%
Wana02}.

For case~a (Fig.~13a), the predicted area in which stars are detected
(shown by the grey image) is small and close to the average stellar
abundance (thick-solid line) and the ISM value (thick-dotted line). Most
of the observed stars distribute out of its 90\% confidence interval
(thin-solid line). This is due to a weak dependence of the Eu/Fe value
on the progenitor mass, since all SNe are assumed to be the $r$-process
site. In fact, this is rather similar to the observed abundances of
$\alpha$ elements with exceedingly small dispersion (e.g., Mg), which
have only mild dependence of the yields on the progenitor mass. In
contrast, large dispersions are predicted for cases~b-d. This is
explained as follows. The star formed by the SN that undergoes
$r$-process inherits the large amount of Eu. This results in the higher
[Eu/Fe] than the ISM value. On the other hand, the star formed by the SN
without $r$-process but with Fe ejecta has the [Eu/Fe] value below the
ISM line. As a result, a large dispersion of the [Eu/Fe] values appears. 
The dispersion converges as the metallicity increases, which also can be
seen in the observed stars, since the formed stars in the
high-metallicity ISM are less affected by the individual SNe.

Figs.~13a-d clearly demonstrate that the limited mass range ($\sim 10\%$
of all SN events) of the progenitor stars that undergo the $r$-process
naturally explain the observed large star-to-star scatters of the
$r$-process elements relative to iron. In addition, a small (or no) iron
production strengthens the dispersion owing to the appearance of stars
with high [Eu/Fe] values, which can be seen in case~b. A significant
difference among these three cases appears, however, in the areas with
the sub-solar [Eu/Fe] ($< 0$) values near [Fe/H] $= -3$. Stars with low
[Eu/Fe] values must appear if the $r$-process elements originate from
SNe near their lower-mass end (Fig.~13b) owing to their delayed
appearances, as also predicted by an earlier work \cite{Math92}. On the
other hand, few stars are expected to appear in this region for case~d,
since the ISM is initially enriched with the $r$-process elements by
massive stars. Case~c with the intermediate mass range of SNe lies
between cases~b and d. Most of the observed stars, in particular when
including those with the lowest [Eu/Fe] values at [Fe/H] $\sim -3$
(large double circles in Fig.~13), locate near the average value (thick
line) predicted in case~b, distributing within the 50\% confidence
interval (Fig.~13b). For case~c, most of these stars distribute below
the average line, but still within the 90\% confidence interval
(Fig.~13c). For case~d, many observed stars are out of the 90\%
confidence interval.

This might support the SNe near their lower-mass end, e.g., collapsing
O-Ne-Mg cores from $8-10 M_\odot$ stars (\S~2.2), as the astrophysical
$r$-process site, although its explosion mechanism (i.e., ``prompt'' or
``delayed'') cannot be constrained. The little production of $\alpha$
and iron-peak elements in this site \cite{Qian03, Wana03} is also
consistent with the observed small scatters of these elements in
extremely metal-poor stars \cite{Cayr04}, by adding only $r$-process
elements to the formed star. It should be noted that the slightly
shifted mass range, e.g., $10-11 M_\odot$, which corresponds to the SNe
from collapsing {\it iron} cores near their low-mass end with the
relatively small ejection of $\alpha$ and iron-peak elements, would
result in a similar outcome. The SNe from more massive progenitors,
e.g., $20-25 M_\odot$ (case~c), as proposed to be a possible case in the
neutrino wind scenario (\S~2.1.4), cannot be excluded either, with the
current limited number of stars having the measured Eu values. Further
detections of Eu in the stars at [Fe/H] $< -3$ without any selection
biases, are highly desired. On the other hand, the SNe near their high
mass end, which may include ``pair-instability SNe'' or ``collapsars'',
are less likely to be the origin of heavy $r$-process nuclei. It is
interesting to note that the large $r$-processed material per event
owing to the limited mass range would increase the chance of direct
detection of $r$-process elements in nearby SN remnants by future
observations. In particular, detection of gamma-ray emission from the
decay of $r$-process nuclei would prove that the SNe with certain masses
are the $r$-process site \cite{Qian98, Wana03}.

It should be cautioned that the ISM is assumed to be {\it homogeneous}
in the current models, which must be {\it inhomogeneous} to some extent
at the early Galactic history. The chemical evolution study with a fully
inhomogeneous ISM model shows that the stars with low [Eu/Fe] values
(such as in Fig.~13b) always appear as far as the SNe that undergo
$r$-processing are restricted to a certain progenitor mass range, even
if the range is assumed to the high-mass end \cite[$20-50 M_\odot$, see
Fig.~3 in][]{Arga04}. On the other hand, such a model results in large
star-to-star scatters of other elements, e.g., [$\alpha$/Fe]
\cite{Arga00}, which conflicts with the recent spectroscopic studies of
extremely metal-poor stars \cite[e.g.,][]{Cayr04}. This might imply that
the ISM at the early Galaxy was efficiently mixed (i.e., close to the
homogeneous ISM), as far as in the region where stars formed. Future
comprehensive studies of {\it inhomogeneous} Galactic chemical evolution
that account for {\it both} the large scatters of [$r$/Fe] and the small
scatters of [$\alpha$/Fe] observed in extremely metal-poor stars
\cite[e.g.,][]{Ishi03, Karl05a, Karl05b}, as well as more measurements
of neutron-capture elements in stars with [Fe/H] $< -3$, will be needed
before drawing any firm conclusions.

If ``NS mergers'' instead of SNe were taken as the major $r$-process
site in the current chemical evolution model, the result would be in
disagreement with the observed stellar abundances, as examined in a
recent work \cite{Arga04}. The reason is that the expected small event
rate $\sim 10^{-5}$~yr$^{-1}$ \cite{vand96} (i.e., the large $r$-process
amount per event to be the dominant $r$-process origin) with the long
period needed for a coalescence results in an extremely large scatter of
[Eu/Fe] as well as a significant delay of its appearance. The same may
hold for AICs, whose event rate is estimated to be similar
\cite{Bail90}. A word of caution is, however, required concerning the
treatment of NS mergers here. The Galactic evolution of NS mergers (or
AICs) as well as the nature of their remnants are highly uncertain. In
addition, the NS mergers may not induce star formation as assumed for
SNe owing to their smaller kinetic energy, and thus not necessarily lead
to a large scatter of the $r$-process elements in stars. A lack of
$\alpha$ and iron-peak elements in the ejecta of NS mergers
\cite{Frei99b, Gori05a} (and of AICs, perhaps) with their uncertain
Galactic evolution makes it quite difficult to determine {\it when} (or,
at which metallicity) the observed stars received the $r$-process
material from the remnants of NS mergers. Obviously, more studies are
needed.

\subsection{``Weak'' $r$-Process}

Besides highly $r$-process-enhanced stars as described in \S~3.1, there
are a significant number of stars (at [Fe/H] $\sim -3$) that show
enhancements of {\it only} light $r$-process nuclei such as Sr, Y, and
Zr \cite{John02b, Ishi05}. In particular, a large dispersion has been
found in [Sr/Ba] at low metallicity \cite{Ryan96, John02b, Hond04},
suggesting that the lighter elements such as Sr have a different origin
from the ``main'' $r$-process that produces Ba and heavier
elements. This may be interpreted as a result of ``weak'' (or failed)
$r$-processing with insufficient (or no) free neutrons at the beginning
of $r$-process, in which only light $r$-process nuclei are produced as
can be seen in Figs.~3a, 4, and 8a-b. HD~122563 \cite{Hond05} may be one
of such stars that have abundance distribution of the weak $r$-process
(Fig.~14). However, the boundary mass number that divides this ``weak''
$r$-process and the ``main'' $r$-process has been unknown, owing to a
limited number of stars that have the measured abundances located
between the first and second $r$-process peaks, e.g., Ru, Rh, Pd, and
Ag. In the following, our recent results of Galactic chemical evolution
are presented \cite{Ishi05}, which may enable us to determine the
typical boundary of these two $r$-processes.

\begin{figure}[t]
\begin{minipage}[t]{68mm}
\includegraphics*[width=\textwidth]{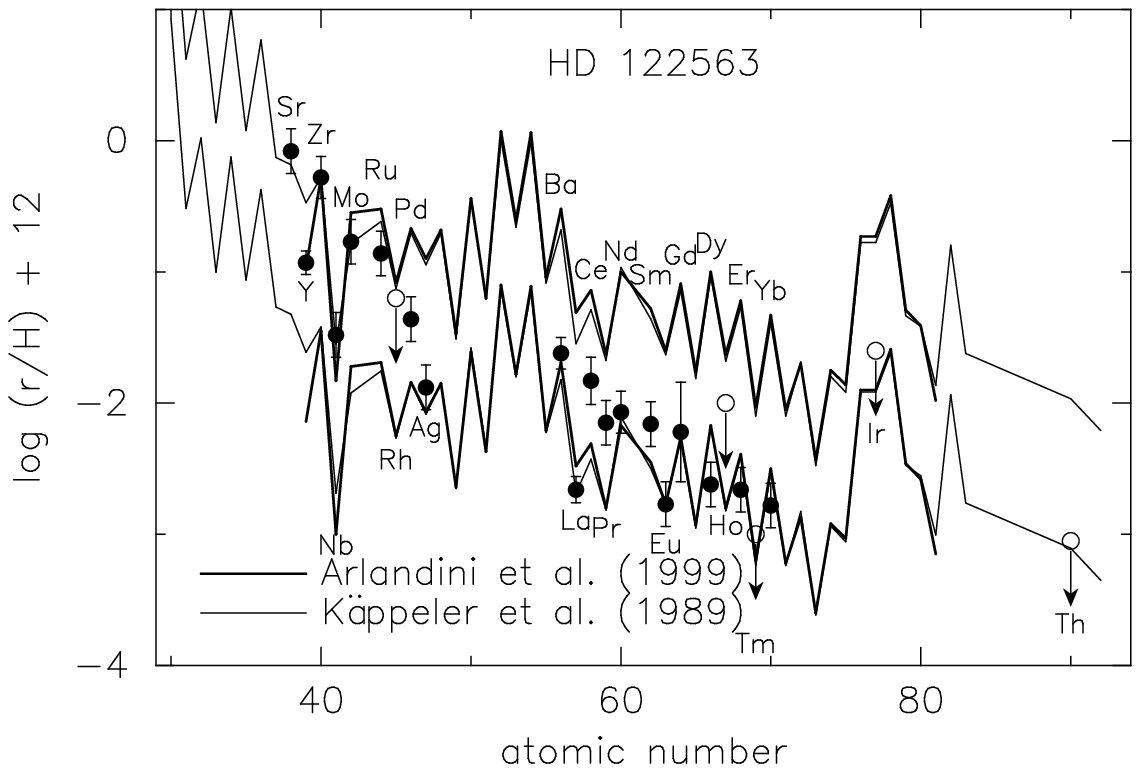} \caption{Same as Fig.~12,
but for HD~122563 \cite{Hond05}. The metallicity of this star is [Fe/H]
$= - 2.7$. The solar $r$-process abundances are scaled to match the
observed Zr and Eu abundances.}
\end{minipage}
\hspace{\fill}
\begin{minipage}[t]{68mm}
\includegraphics*[width=\textwidth]{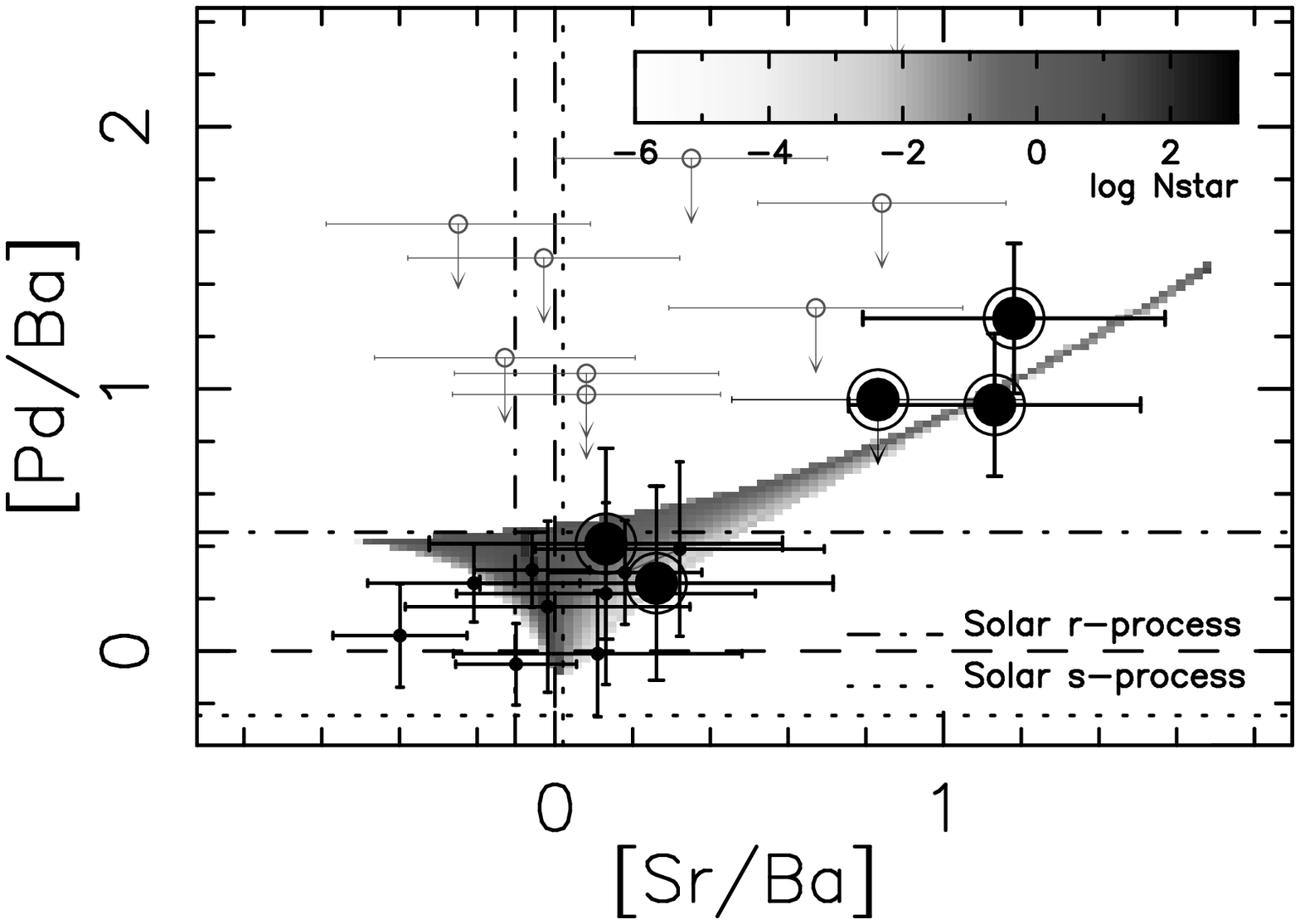} \caption{Same as Fig.~13,
but for [Pd/Ba] as a function of [Sr/Ba]. The large double circles
indicate the stars obtained from our recent observations with SUBARU/HDS
\cite{Ishi05}.}
\end{minipage}
\end{figure}

As representative of the weak and main $r$-processes and their
intermediate, Sr, Ba\footnote{There are few ``weak'' $r$-process stars
with the measured values of Eu, but of Ba.}, and Pd are taken here,
respectively. Fig.~15 shows the [Pd/Ba] values as a function of [Sr/Ba]
for extremely metal-poor stars ($-3.1 \le [{\rm Fe}/{\rm H}] \le -2.3$). 
Note that for stars in this metallicity range, no $s$-process
contribution to Ba abundances are thought to occur, which is consistent
with their mostly pure solar $r$-process ratios of Ba/Eu. The stellar
sample is taken from the recent literature \cite{Hill02, John02b,%
Sned03, Hond04}, along with the five stars (large double circles)
obtained from our recent observations with SUBARU/HDS \cite{Ishi05}. By
definition, the [Sr/Ba] value increases with the contribution of weak
$r$-process to the stellar abundances, when compared to the solar
$r$-process value (dot-dashed line). The [Pd/Ba] value would increase
linearly with [Sr/Ba] (with the slope of unity) if the Pd abundances
were in proportion to those of Sr. On the other hand, [Pd/Ba] would show
no increase if the Pd abundances were in proportion to those of Ba. The
observed stars show a mild correlation with the slope less than unity,
indicating Pd originates from {\it both} the weak and main
$r$-processes.

The chemical evolution model described in \S~3.2 \cite{Ishi99} is used
to determine the fractions of the weak r-process component to the {\it
total} (i.e., both the weak and main $r$-processes) production of Sr,
Pd, and Ba. Here, the weak and main $r$-process sites are assumed to be
the stars of $8-10 M_\odot$ and $20-30 M_\odot$, respectively
\cite[changes of these mass ranges do not affect the result
qualitatively, see][]{Ishi05}. A reasonable fit to the observed stars
can be obtained when the fractions of the weak $r$-process component to
Sr, Pd, and Ba are assumed to be 60\%, 10\%, and 1\%, respectively
(Fig.~15). Thus, the {\it typical} weak $r$-process may synthesize
mainly lighter nuclei up to $A \sim 100$ ($Z \sim 40$). However the
number of the Sr-rich stars with the measured Pd abundances is so small
(two measured values and one hard upper limit) that more observations
will be needed before drawing the conclusion. This boundary mass
estimated here is significantly smaller than $A \sim 140$ that is
suggested by the meteoritic analysis \cite{Wass96}. It should be
emphasized that the main $r$-process may produce {\it all} the
$r$-process nuclei with the solar $r$-process-like distribution but,
perhaps, with slightly underabundant lighter nuclei when subtracting the
weak $r$-process component. As can be seen in Fig.~15, there is
currently no observed star that shows strong excess of {\it only} heavy
$r$-process elements with $A > 130$ (i.e., with significantly low
[Sr/Ba] and [Pd/Ba] values compared to the solar $r$-process ratios) as
suggested in previous studies \cite{Wass96, Frei99b}.

\subsection{Cosmochronology}

A few actinides such as $^{232}$Th and $^{238}$U are regarded as
potentially useful cosmochronometers because their long radioactive
decay half-lives ($^{232}$Th: 14.05~Gyr; $^{238}$U: 4.468~Gyr) are
significant fractions of the expected age of the universe. The excellent
agreement of the relative abundances of neutron-capture elements in
CS~22892-052 with the solar $r$-process pattern (\S~3.1) initially
suggested that Th might serve as a precise cosmochronometer
\cite{Sned96, Cowa97, John01}. The time that has passed since the
production of Th observed now in the atmosphere of such an old halo star
can be regarded as a hard lower limit on the age of the universe. One
advantage of the actinide chronology is that, once the initial and
current values of an actinide relative to an stable $r$-process element
($r$), e.g., Eu, in the star are specified, the age of the star depends
only on the half-life of its actinide determined in the laboratory. That
is, one is not forced to invoke complex models of Galactic chemical
evolution, which no doubt involve large uncertainties in themselves.

The {\it initial} production ratio of Th/$r$ has been usually determined
by fitting theoretical nucleosynthesis results to the solar $r$-process
pattern, with the assumption that the $r$-process pattern is {\it
universal} over the actinide region in all astrophysical environments
\cite[e.g.,][]{Cowa99, Gori99}. However, there are an increasing number
of evidences that the universality does not hold for actinides by the
discoveries of Th-rich halo stars \cite{Hill02, Hond04, Yush05} whose
Th/Eu values are higher than that of the solar $r$-process ratio. These
{\it old} halo stars would be younger than the solar system if the
initial Th/Eu were taken to be universal. Theoretical studies of
$r$-process calculations also support the {\it non-universality} of the
$r$-process abundances beyond the Pt-peak nuclei ($A \sim 200$), as can
be seen in Fig.~8 \cite[see also][]{Wana02}. Therefore, any age
estimates that demand the universality of the $r$-process pattern may in
fact be unreliable.

\begin{figure}[t]
\begin{minipage}[t]{68mm}
\includegraphics*[width=\textwidth]{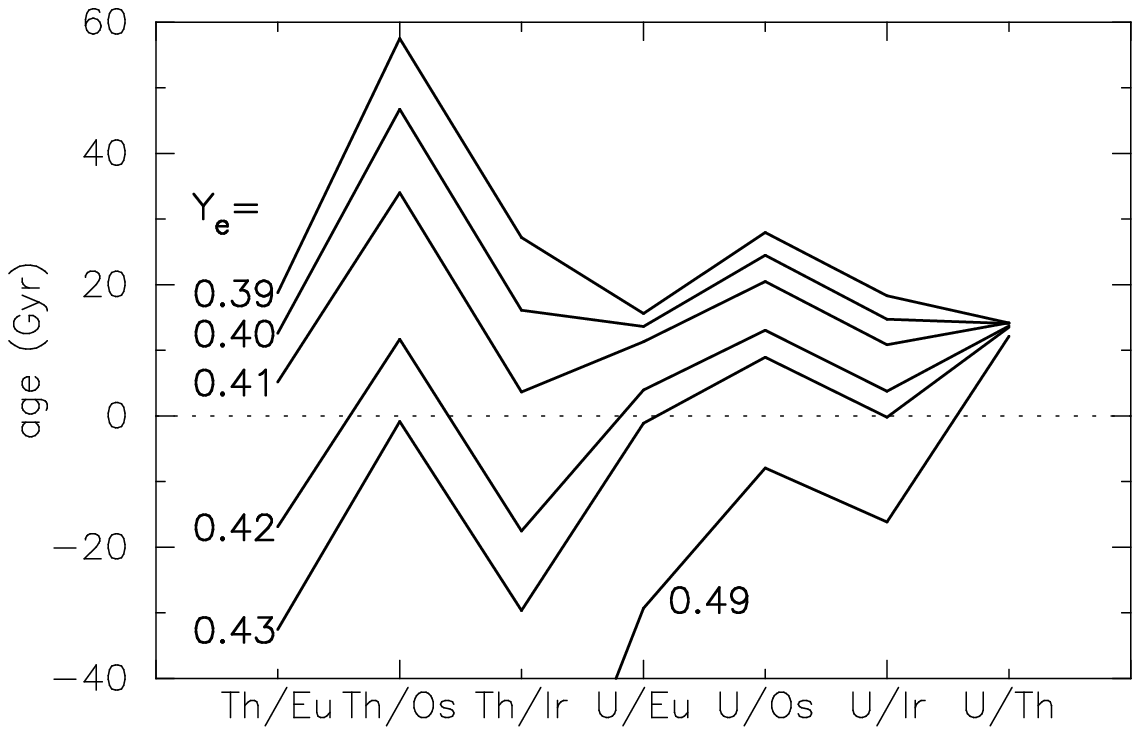} \caption{Ages of
CS~31082-001 derived from various chronometer pairs, comparing the
abundances obtained from the spectroscopic analysis \cite{Hill02} and
the theoretical estimate \cite{Wana02} based on the neutrino wind model
as described in \S~2.1. The robustness of the U-Th pair is clearly
shown. The superiority of the U-$r$ pairs compared to those of Th-$r$
can also be seen.}
\end{minipage}
\hspace{\fill}
\begin{minipage}[t]{68mm}
\includegraphics*[width=\textwidth]{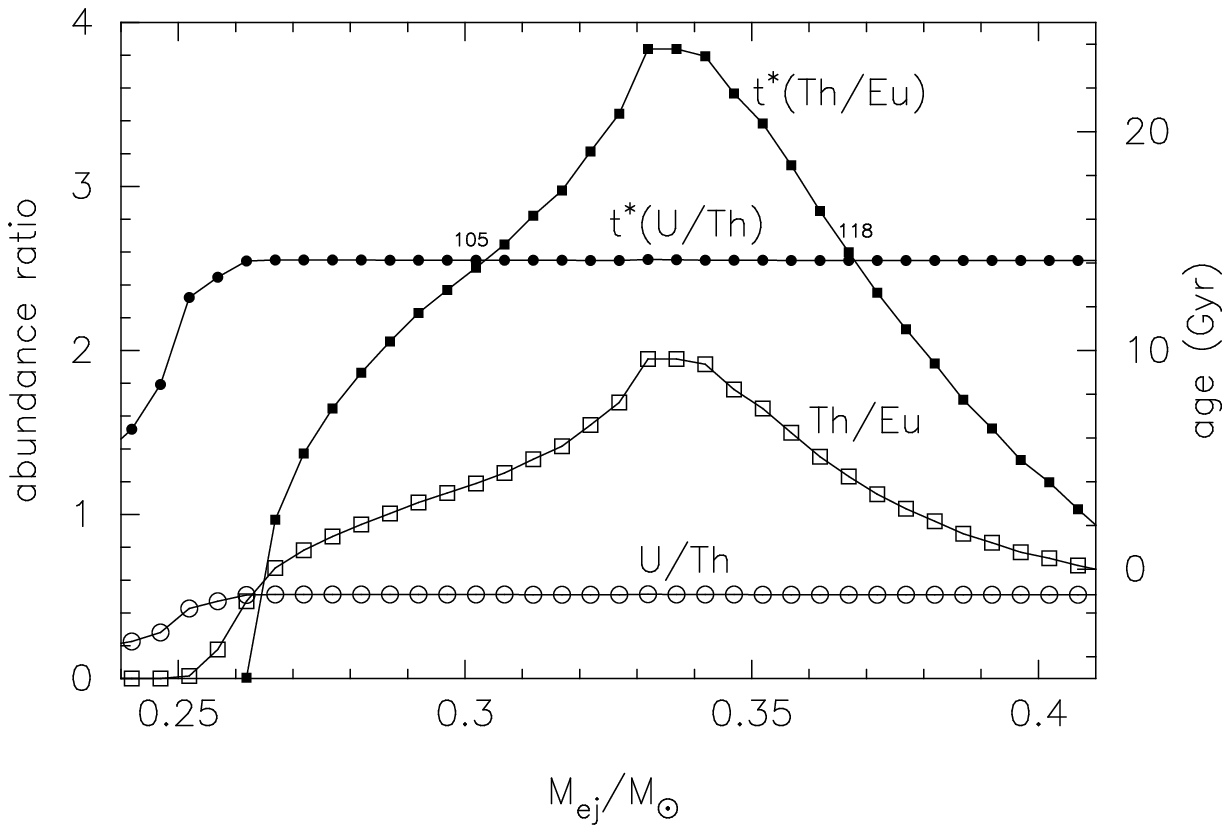} \caption{Mass-integrated
abundance ratios Th/Eu (open squares) and U/Th (open circles) from the
surface of the core to the mass point $M_{\rm ej}$ in the prompt
explosion model \cite{Wana03} as described in \S~2.2. The surface of the
O-Ne-Mg core is at mass coordinate zero. Ages of CS~31082-001, $t^*
({\rm Th/Eu})$ (filled squares) and $t^* ({\rm U/Th})$ (filled circles)
inferred by these ratios, are also shown.}
\end{minipage}
\end{figure}

The discovery of the second highly $r$-process-enhanced halo star
CS~31082-001 \cite{Cayr01, Hill02} has provided a powerful new tool for
age determination by virtue of the detection of uranium \cite{Gori01,%
Wana02, Scha02}. Because the half-life of $^{238}$U is one-third that of
$^{232}$Th, uranium can, in principle, provide a far more precise
cosmochronometer than thorium\footnote{$^{235}$U may have mostly
$\alpha$ decayed away because of its relatively short half-life
(0.704~Gyr).}. Furthermore, we are able to determine the initial
$r$-process abundance curve in a star with the constraint that the ages
derived from {\it both} the ratios Th/$r$ and U/$r$ (or U/Th) provide
the same value. This ``U-Th cosmochronology'' that does not assume the
universality of the $r$-process abundance pattern is a far more reliable
age-dating technique than that with solely Th-$r$. Fig.~16 shows the
ages of CS~31082-001 derived from various chronometer pairs, comparing
the abundances obtained from the spectroscopic analysis \cite{Hill02}
and the theoretical estimate \cite{Wana02} based on the neutrino wind
model ($M = 2.0 M_\odot$ and $R_\nu = 10$~km) as described in
\S~2.1. Here, $Y_e$ is taken to be a free parameter, which can be
constrained to be $\approx 0.40$ so as the ages obtained from, e.g.,
Th/Eu and U/Th give the same value. Fig.~17 shows the ages derived from
the ratios Th/Eu and U/Th based on the prompt explosion model of a
collapsing O-Ne-Mg core as described in \S~2.2, as functions of the
ejecta mass $M_{\rm ej}$. The free parameter here, $M_{\rm ej}$, can be
constrained to be $0.30 M_\odot$ or $0.37 M_\odot$, where both ages from
the ratios Th/Eu and U/Th give the same value. Interestingly, the ages
of CS~31082-001 derived from the above two different astrophysical
scenarios are the same -- $14.1 \pm 2.4$~Gyr (the error only includes
that arising from the observations). This demonstrates that chronometric
estimates obtained using the U-Th pair are mostly independent of the
astrophysical conditions considered, since these species separated by
only two units in atomic number.

As far as the U-Th pair is concerned, therefore, the ingredients of
nuclear data \cite{Gori01} as well as the estimated observational
errors, rather than the $r$-process site, are crucial for the age
determination. Note that the replacement of the nuclear mass formula
(see \S~2.4) from DM applied here \cite{Hilf76} to FRDM \cite{Moel95} or
the recent microscopic models \cite[HFB-2 and HFB-7,][]{Gori02, Samy03,
Gori03} results in only small changes in age of the star
($14.0-14.3$~Gyr). The uncertainties in fission reactions may not affect
the age significantly as far as the contribution of the fission fragment
is not dominated in the final $r$-process abundances \cite{Gori01,%
Wana02, Wana03}. It should be noted that there is a serious problem,
that is, the measured Pb abundance is substantially lower than the scaled
solar value as can be seen in Fig.~12 \cite{Plez04} and also than the
theoretical estimates \cite{Wana03}. Improved observational
determination of the U/Th ratio in CS~31082-001 as well as the
measurement of Bi (in addition to Pb) that is produced mainly by
$\alpha$-decay from actinides, and the identification of a greater
number of highly $r$-process-enhanced, metal-poor stars, will be
obviously needed before regarding the age-dating technique with the U-Th
pair considered here to be confident.

\section{Conclusions}

The astrophysical $r$-process site is still unknown. Recent theoretical
works of $r$-process calculations suggest some scenarios, in particular
the ``neutrino wind'' or the ``prompt explosion'' arising from the core
collapses of massive stars, or the ``NS merger'' to be promising,
although all these involve severe problems that remain to be solved. On
the other hand, recent Galactic chemical evolution studies as well as
spectroscopic studies of extremely metal-poor halo stars imply the
$r$-process origin to be the core-collapse SNe from the restricted
progenitor-mass range (possibly near their lower-mass end). In addition,
the presence of the ``weak'' $r$-process that produces only lighter
$r$-process nuclei is suggested, while the ``main'' $r$-process may
produce all $r$-process nuclei up to the actinides species. All these
studies of $r$-process calculations and Galactic chemical evolution in
the last decade have shown remarkable progresses toward better
understanding of the $r$-process, and the ongoing works will shed light
on this long-standing mystery.

\section*{Acknowledgements}

This work was supported in part by a Grant-in-Aid for the Japan-France
Integrated Action Program (SAKURA), awarded by the Japan Society for the
Promotion of Science, and Scientific Research (17740108) from the
Ministry of Education, Culture, Sports, Science, and Technology of
Japan.

\end{document}